\documentclass[useAMS,usenatbib]{mn2e}

\usepackage{lscape}
\usepackage{graphicx}

\title[Empirical isochrones]{Empirical isochrones and relative ages
  for young stars, and the radiative-convective gap} \author[N.J.
Mayne et al.]{N.J. Mayne$^{1}$\thanks{E-mail: nathan@astro.ex.ac.uk
    (NJM)}, Tim Naylor$^{1}$, S. P.  Littlefair$^{2}$, Eric S.
  Saunders$^{1}$ and R.  D. Jeffries$^{3}$\\
  $^{1}$ School of Physics,
  University of Exeter, Stocker Road, Exeter, EX4 4QL.\\
  $^{2}$ Department of Physics and Astronomy, University of Sheffield.\\
  $^{3}$ Astrophysics Group, School of Physical and Geographical
  Sciences, Keele University, Keele, Staffordshire ST5 5BG}

\begin{document}

\date{Accepted ?. Received ?; in
  original form ?}

\pagerange{\pageref{firstpage}--\pageref{lastpage}} \pubyear{2006}

\maketitle

\label{firstpage}

\begin{abstract}  
  We have selected pre-main-sequence stars in 12 groups of notional 
  ages ranging from 1 Myr to 35 Myrs, using heterogeneous
  membership criteria.
  Using these members we have constructed empirical isochrones in
  \textit{V}, \textit{V-I} colour magnitude diagrams (CMDs).
  This allows us to identify clearly the gap between the radiative
  main sequence and the convective pre-main-sequence (the R-C gap). 
  We follow the evolution of this gap with age and show that it can be
  a useful age indicator for groups less than $\simeq15$ Myrs old.
  We also observe a reduction in absolute spreads about the sequences
  with age.
  Finally the empirical isochrones allow us to place the groups in
  order of age, independently of theory.
  The youngest groups can be collated into three sets of similar ages.
  The youngest set is the ONC, NGC6530 and IC5146 (nominally 1 Myrs);
  next Cep OB3b, NGC2362, $\lambda$ Ori and NGC2264 (nominally 3
  Myrs); and  finally $\sigma$ Ori and IC348 (nominally 4-5 Myrs).
  This suggests Cep OB3b is younger than previously thought, and IC348 
  older.
  For IC348 the stellar rotation rate distribution and fraction of 
  stars with discs imply a younger age than we derive.
  We suggest this is because of the absence of O-stars in this
  cluster, whose winds and/or ionising radiation may be an important
  factor in the removal of discs in other clusters.
\end{abstract}

\begin{keywords}
  stars:evolution -- stars:formation -- stars: pre-main-sequence --
  techniques: photometric -- catalogues -- (stars) Hertzsprung-Russell
  H-R diagram
\end{keywords}

\section{Introduction}
Most ages for stellar groups, sub-groups, clusters or associations are
derived from fitting observations to theoretical isochrones. It is 
well known that these models often deviate systematically from the
data.
For instance if the isochrone fits part
of a sequence it may systematically deviate from the observed sequence
in other sections \citep[e.g.][]{2004A&A...415..571B, 2004ApJ...600..946P}. In
addition if a sequence of stars is fitted to an isochrone in one
colour, derived parameters such as the age may be different from those
derived in another colour \citep[e.g.][]{2002MNRAS.335..291N}.  In this paper
we create empirical
isochrones.  These have the advantage of necessarily fitting the
entire data sequence, allowing us to create a relative age ladder. We
can then derive ages for fields where different regions of a sequence
are available by selecting appropriate fiducial sequences covering the
required colour range. The obvious disadvantage of our empirical
isochrones is that they are unable to provide absolute age
information.  
Although this is the first attempt to create
such a ladder, carried out with data that uses slightly different photometric
systems and heterogeneous selection criteria, the method is clearly 
effective.

As we have used a range of ages to create the empirical
isochrones we are able to examine how these isochrones change with
age. In many fields this reveals a clear gap between the convective
lower-mass pre-main-sequence (PMS) and the radiative stars close to or
on the main sequence (MS). We refer to this feature as the
radiative-convective gap (R-C gap).

For this study we have used the optical bands \textit{V} and
\textit{I}.  This enables us to minimise the effect of accretion,
disc presence and chromospheric activity on our empirical isochrones.
Accretion will generally affect the flux short-ward of the \textit{B}
band \citep{1998ASPC..154.1709G} with disc emission important at
longer wavelengths, first becoming significant in the \textit{H} and
\textit{K} bands \citep{1998apsf.book.....H}. Lastly,
\cite{2003AJ....126..833S} have shown that magnetically generated
chromospheric activity in young stars primarily causes perturbations
and scatter in the \textit{B-V} and \textit{U-B} colours, whereas the
\textit{V}, \textit{V-I} CMD is unaffected.

Our paper is laid out as follows.
The data collection and reduction are detailed in Section
\ref{reduction}. 
To create the empirical isochrones we select members in each field
using a variety of criteria (see Appendix \ref{Known_memberships} and
Section \ref{Positional_isolation}) and fit a curve through these in a
\textit{V}, \textit{V-I} CMD (see Section \ref{fitting}). We then
apply a distance modulus and correct for extinction (see Section
\ref{Parameters}). 
The reader may
wish to move straight to Section \ref{Data_comparison}.  Here we
present the data corrected for distance and extinction, allowing
comparisons to be made with theory and also between sequences from
clusters, groups or associations of different ages. 
Finally we combine
the sequences to form a fiducial relative age ladder (see Section
\ref{Relative_ages}), which we then use 
to find new age estimates
for six fields (see Section \ref{ladder_use}).  In Section
\ref{conclusion} we summarise our conclusions. Appendix
\ref{Known_memberships} contains a discussion of the literature
membership data, while in Appendix \ref{appendix2} we
discuss the distances and extinctions from the literature that we have
used.

\section[]{DATA COLLECTION}
\label{reduction}
The dataset used in this paper is from both new observations and
literature sources.

\subsection{New observations}
The images for the majority of this dataset were obtained in
\textit{BVI} with the 2.5m Isaac Newton Telescope (INT), situated on
La Palma, equipped with the four EEV CCD Wide Field Camera (WFC). The
filters used were Sloan \textit{i'} and Harris \textit{V} and
\textit{B}. The datasets used were collected on the nights of the 27th
September and the 5th of October 2004, when the range of seeing was
1-2'' and 1-1.5'' on each night respectively. Both nights were
photometric. The observational details are given in Table
\ref{observe}. Standards from \cite{1992AJ....104..340L} were observed
on these photometric nights, tying our calibration to
\textit{B},\textit{V} and \textit{$I_c$} Landolt standards.

\begin{table*}
\begin{tabular}{|l|l|l|}
Field&Filter&Exposure time (secs) ($\times$1 unless stated)\\
\hline
h Per&\textit{I}&3 and 30\\
&\textit{V}&1, 10 and 100\\
&\textit{B}&2 and 20\\
\hline
$\chi$ Per&\textit{I}&3 and 30\\
&\textit{V}&1, 10 and 100\\
&\textit{B}&2 and 20\\
\hline
NGC7160&\textit{I}&3 and 30\\
&\textit{V}&1, 10 and 100\\
&\textit{B}&2 and 20\\
\hline
NGC2264 (4 fields)&\textit{I}&2 and 20\\
&\textit{V}&2 and 20\\
&\textit{B}&4 and 40\\
\hline
IC348&\textit{I}&2 and 20\\
&\textit{V}&2 and 20\\
&\textit{B}&4 and 40\\
\hline
Cep OB3b&\textit{I}&5($\times$2), 30($\times$2) and 300($\times$2)\\
&\textit{V}&5, 30 and 350($\times$4)\\
\hline
$\sigma$ Ori (Fields 1,2 and 4)&\textit{I}&2 and 20\\
&\textit{V}&2, 10, 100 and 350($\times$4)\\
$\sigma$ Ori (Field 3)&\textit{I}&2 and 20\\
&\textit{V}&2, 10, 100 and 350\\
&\textit{B}&4\\
\hline
\end{tabular}
\caption{Exposure times for each field.\label{observe}}
\end{table*}
\begin{table*}
\begin{tabular}{|l|l|l|l|}
\hline
Region&Table Number&Data&Source (as table \ref{Sources})\\
\hline
NGC2547&4&Members&1 (X-ray)\\
&5&Members&2 (Spectroscopy)\\
\hline
NGC7160&6&Full catalogue&\\
&7&Members&3 (Extinction)\\
\hline
NGC2264&8&Full catalogue&\\
&9&Members&6 (Periodic Variables)\\
&10&Members&4 (X-ray)\\
&11&Members&5 (H$\alpha$)\\
&12&Members&26 (Proper Motion)\\
\hline
Cep OB3b&13&Full catalogue&\\
&14&Members&7 (X-ray)\\
&15&Members&8 (Spectroscopy)\\
&16&Members&18 (Periodic Variables)\\
&17&Members& 19 (H$\alpha$)\\
&18&Members&25 (X-ray)\\
\hline
$\sigma$ Ori&19&Full catalogue&\\
&20&Members&11 (X-ray)\\
&21&Members&12 (Spectroscopy)\\
\hline
IC348&22&Full catalogue&\\
&23&Members&10 (Periodic Variables)\\
&24&Members&16 (X-ray)\\
&25&Members&15 (X-ray)\\
&26&Members&9 (Spectroscopy)\\
&27&Members&17 (H$\alpha$)\\
\hline
h and $\chi$ Per&28& Full catalogue (combined)\\
&29&Positionally isolated&\\
\hline
\end{tabular}
\caption{The catalogues presented in this paper.\label{table}}
\end{table*}

\subsection{Data reduction}
The images were debiased, and then flat fielded using median stacks of
many sky flat frames. The \textit{i} band images were then defringed
using a library fringe frame.  Known bad pixels were flagged. The
data, including the standards, were then reduced following the optimal
extraction algorithm detailed in \cite{1998MNRAS.296..339N} and
\cite{2002MNRAS.335..291N}.  We used the 2 Micron All Sky Survey
(2MASS) to provide an astrometric solution for each field accurate to
$\simeq$0.1 arcsec. The reduction process provided catalogues of
photometry with flags and uncertainties. The flags are explained in
\cite{2003MNRAS.346.1143B}, excepting the flag T. These are objects
were the profile correction \citep[see][]{2002MNRAS.335..291N} fails due to a paucity of stars in the
field. Therefore aperture photometry is performed using an aperture of
the same radius as that to which the optimally extracted fluxes are
corrected.
Since aperture photometry is noisier than optimal photometry, we only
carry out this proceedure for the brighter stars, hence the fluxes for
fainter stars in the same image will be optimally extracted, and
flagged as having poor profile corrections (H).

Photometric calibration coefficients were calculated from the
standard star observations using a similar procedure to that detailed in
\cite{2003MNRAS.341..805P}.
For each colour or magnitude we allow the following free parameters:
one extinction coeffficient per night; one colour term for each CCD;
one zero point for each night, with offsets with respect to this
(which do not change from night to night) for three of the four CCDs.
The standards used had colour ranges of $-0.30<\textit{V-I}<2.87$ and
$-0.30<\textit{V-I}<2.85$, with airmass ranges of $1.13-1.32$ and
$1.14-1.22$, on September 27th and October 5th 2004 respectively. 

We estimate the combined
uncertainty in profile correction and transformation to the Landolt
system by adding a magnitude independent uncertainty.
Adding 0.02 mags in each band gave a reduced $\chi^2$ of approximately
one in \textit{V} and \textit{B-V}.
The value of reduced $\chi^2$ was somewhat higher in \textit{V-I}
(1.4), which we suspect is due to the poor match between our $i$
filter and those used by Landolt.

In the cases where multiple fields were observed (NGC2264, $\sigma$
Ori and h and $\chi$ Per) the catalogues were combined using the
method described in \cite{2002MNRAS.335..291N}. Overlapping regions
were used to normalise the photometry to provide a combined catalogue.
As detailed in \cite{2002MNRAS.335..291N} the RMS of the overlap
region is a good measure of the accuracy of the profile correction.
This suggests a magnitude independent uncertainty of approximately
0.01 mags which should be added to all bands when comparing objects
well separated on the CCDs. Using the overlap region suggest an
accuracy of 1\% in both \textit{V} and \textit{B-V} and around 2\% in
\textit{V-I}. We believe the higher uncertainty in \textit{V-I} is
caused by a lack of redder standards across all the CCDs. This
uncertainty is not included in the catalogues.  For all subsequent
analysis the combined catalogues have been used.

All of the new catalogues used for this paper are freely available from the
cluster collaboration home
page\footnote{http://www.astro.ex.ac.uk/people/timn/Catalogues/description.html},
and the CDS archive (sample can be seen as Table \ref{catalogues}).
For membership selection and CMD plots we exclude stars with uncertainties
greater than 0.1 mags in colour or magnitude. The machine readable
catalogues retain all the data.

To aid navigation through this paper an index to the catalogues is
included as Table \ref{table}. Full catalogues are presented for each
field with new photometry in addition to catalogues of stars which
fulfill given membership criteria.

\subsection{Literature data}
\label{literature_data}
\begin{figure}
  \vspace*{174pt}
  \includegraphics{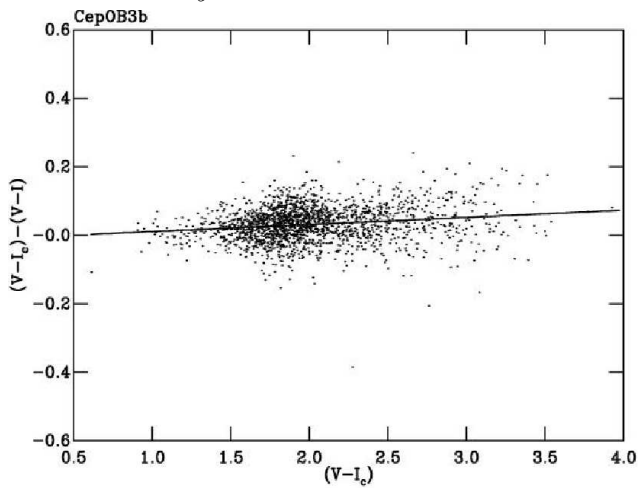}
  \caption{The difference in $V-I$ for Cep OB3b between data taken 
    in the system used here and using Cousins $I$ filter.
  The line is the best fit through the data.
\label{cous}}
\end{figure}

Literature observations were taken in all cases (except where stated)
in the Johnson-Cousins photometric system using $\textit{I}_c$
(Cousins \textit{I}). This raises the question of what effect the use
of two different \textit{I} filters has on our results. The difference
is illustrated in Figure \ref{cous}. The data for Sloan \textit{i},
for stars in Cep OB3b is from this work and the data for Cousins
$\textit{I}_c$, for the same stars is taken from
\cite{2003MNRAS.341..805P}. The symbol \textit{I} from hereon will be
used to represent data taken using a Sloan \textit{i} (\textit{i'})
filter tied to the Landolt standards, with $\textit{I}_c$ being data
taken in Cousins $\textit{I}_c$ tied to Landolt standards. The
difference (\textit{V-I$_c$})-(\textit{V-I}) is plotted as a function
of (\textit{V-I$_c$}) in Figure \ref{cous}. The data can be
represented by the following linear fit: (\textit{V-I}) =
0.979(\textit{V-I$_c$}) + 0.009.  This result shows that two stars
with an apparent \textit{V-I} of 3 mags, one cool and unredenned i.e.
a typical Landolt standard, and another hotter and more heavily
reddened, need not have the same \textit{V-I$_c$}. As demonstrated by
the extreme example of CepOB3b ($A_V\simeq3$ mags), even at a
$\textit{V-I}=3$ the shift is $<0.1$ mags; though note that the
reddest standard in our \textit{V-I} calibration is 2.87. This means
over the colour range covered within this paper the effect is
negligible.

\section[]{SEQUENCE ISOLATION\\*AND FITTING}
\label{isolate_fit}
For each field-of-view (FOV) the CMD includes a high proportion of
background and foreground objects, therefore we faced the common
problem of identifying members.  For most fields we used literature
sources for members. We discuss these fields in Sections
\ref{Seq_members} and \ref{fitting}. For h and $\chi$ Per no
membership data were available for the group, or none at faint enough
magnitudes.  Here we can increase our confidence in selecting the
correct sequence by selecting a limited area of the sky.  This method
will clearly include a higher proportion of non-members.  The field
treated in this way is discussed in Section
\ref{Positional_isolation}.

\subsection{Sequences with memberships}
\label{Seq_members}
In Appendix \ref{Known_memberships} we collect literature membership
data for each field, a summary of which is presented in Table
\ref{Sources}. We use a wide range of membership criteria, each of
which will have an implicit bias. As we are interested in the form of
the sequences in CMD space it is important to be aware of and limit
any bias incurred with respect to the CMD space. X-ray selection will
be biased to weak lined T Tauri stars (WTTS) and binary systems
\citep{2005ApJS..160..390P}. Variability exhibits a bias towards
accreting objects, though periodic variability may exclude many
classical T Tauri stars (CTTS) objects when observed over a long
baseline.  However periodic variability deduced from a short baseline
does not carry this bias \citep{2005MNRAS.358..341L}. The final
frequently used data source was H$\alpha$, which is generally biased
towards CTTS.

There is also a more subtle form of bias due to pre-selection. Because
in many cases surveying all objects in a field is not possible, an
observer will necessarily select a subset of objects. Often this
initial selection is based on the positions of objects in CMD space
thus introducing a bias in CMD space. This is particularly applicable
to spectroscopic data. Where this form of sample selection has
occurred the selection criteria are stated clearly in Appendix
\ref{Known_memberships}.

Samples derived from non-spectroscopic methods of membership selection
have a higher probability of containing contamination from background
(BG) or foreground (FG) objects. This leads to the need for
colour-magnitude selections to remove data points which lie far from
the group sequence. Where we have done this the number of discarded
objects is small relative to the total number of objects and has
little effect on the final sequence (see Section \ref{fitting}). In
each section where such a selection was used, it is shown in the CMD 
and its effect on the result is discussed.

\begin{table*}
\begin{minipage}{140mm}
\begin{tabular}{|l|l|l|l|l|}
\hline
Group&Data Type&Source&Matching&Offsets\\
&&&radius&RA, DEC (arcsec)\\
\hline
NGC2547&X-ray (ROSAT)&1&From 1&0 0\\
&Spectroscopy&2&1 arcsec&0 0\\
\hline
NGC7160&Extinction&3&1 arcsec&0 0\\
\hline
NGC2264&X-ray (ROSAT)&4&6 arcsec&0 0\\
&H$\alpha$&5&1&0 0\\
&Periodic Variables&6&1 arcsec&0 0\\
\hline
Cep OB3b&X ray (ROSAT)&7&From 7 otherwise 14 arcsec&0 0 (both)\\
&X ray (\textit{CHANDRA} ACIS)&25&1&0 0\\
&Spectroscopy&8&1 arcsec (both)&0 0 (both)\\
&H$\alpha$&19&1 arcsec &0 0\\
&Periodic Variables&18&1 arcsec&0 0\\
\hline
$\sigma$ Ori&X-ray (XMM-NEWTON)&11&6 arcsec&0 0\\
&Spectroscopy&12&1 arcsec (both)&0 0 (both)\\
\hline
ONC&Periodic Variables&13&1 arcsec&+1.5 -0.25 and 0 0\\
&X-ray (\textit{CHANDRA})&14&2 arcsec&0 0\\
\hline
IC348&Periodic Variables&10&1 arcsec&-0.1 -0.5\\
&X-ray (\textit{CHANDRA} ACIS)&16&1 arcsec&-0.5 0\\
&X-ray (ROSAT)&15&14 arcsec&0 0\\
&Spectroscopy&9&1 arcsec (both)&0 0 and +0.2 -0.5\\
&H$\alpha$&17&1 arcsec&+0.2 -0.5\\
\hline
NGC2362&H$\alpha$ (Li)&20&Members from 20\\
\hline
$\lambda$ Ori&Spectroscopy (Li)&21&Members from 21\\
\hline
IC5146&H$\alpha$&22&1&0 0\\
\hline
NGC6530&H$\alpha$&23&Members from 23\\
&X-ray (\textit{CHANDRA} ACIS)&24&Members from 24\\
\hline
\end{tabular}
\caption{Table of literature sources and astrometric matching
  criteria. 1 \citet{1998MNRAS.300..331J}, 2
  \citet{2005MNRAS.358...13J}, 3 \citet{2005AJ....130..188S}, 4
  \citet{1999A&A...345..521F}, 5 \citet{2005AJ....129..829D}, 6
  \citet{2004A&A...417..557L}, 7 Second ROSAT PSPC catalogue and
  \citet{1999MNRAS.302..714N}, 8 \citet{2003MNRAS.341..805P} and
  \citet{2002AJ....123.2597O}, 9 \citet{2003ApJ...593.1093L} and
  \citet{1998ApJ...497..736H}, 10 \citet{2004AJ....127.1602C} and
  \citet{2005MNRAS.358..341L}, 11 \citet{2004A&A...421..715S}, 12
  \citet{2005MNRAS.356.1583B} and \citet{2005MNRAS.356...89K}, 13
  \citet{2002A&A...396..513H}, 14 \citet{2003ApJ...582..398F}, 15
  Second ROSAT PSPC catalogue, 16 \citet{2002AJ....123.1613P}, 17
  \citet{1998ApJ...497..736H}, 18 Littlefair et al (in preparation),
  19 \citet{2002AJ....123.2597O}, 20 \citet{2005AJ....130.1805D}, 21
  \citet{2001AJ....121.2124D}, 22 \citet{2002AJ....123..304H}, 23
  \citet{2000AJ....120..333S}, 24 \citet{2005A&A...430..941P}, 25
  \citet{2006ApJS..163..306G} and 26 \citet{1980MNRAS.190..623M}.
  \label{Sources}}
\end{minipage}
\end{table*}

\subsection{Fitting procedure}
\label{fitting}
To explain the stages involved in producing the empirical isochrones
we use an example cluster, NGC2264. This cluster is a good
illustration as it has a large number of members from various sources.

\subsubsection{NGC2264: an example}
\begin{figure}
  \vspace*{174pt}
  \includegraphics{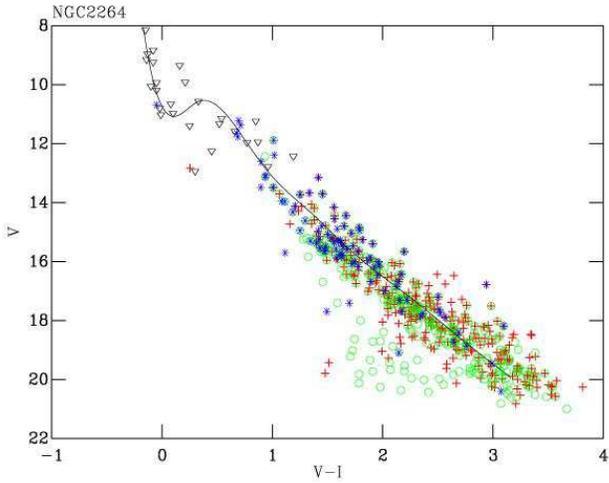}
  \caption{Stars selected as members of NGC2264. 
    Circles are the periodic variables from
    \citet{2004A&A...417..557L}, asterisks are X-ray sources from
    \citet{1999A&A...345..521F}, crosses are H$\alpha$ sources from
    \citet{2005AJ....129..829D} and triangles are proper motion
    members from \citet{1980MNRAS.190..623M}. The spline fit to
    the sequence after a colour-magnitude selection is shown.
    \label{ngc22642}}
\end{figure}

Once the sequence was finalised to stars having properties
indicative of cluster membership (see Appendix \ref{NGC2264_fit}) we
fitted the median filtered sequence with a smooth curve.  The median
filter was used for two reasons. The median is necessarily tied to the
values of real stars. In addition the median allows the effect of
stars lying far from the sequence to be limited, as not all of our
memberships are certain, for instance the X-ray sources in this group.
The median of any sequence was calculated by binning the stars in
\textit{V}-band intervals and assigning the median \textit{V-I} and
\textit{V} values to the median star for that bin.  The bin sizes
were tailored to each field, and we enforced a minimum number of stars
in each bin.  A cubic spline was fitted through the resulting list of
median stars, over the colour range of the stars selected, with the
gradients at the ends tied to a quadratic. The resulting fit for
NGC2264 is shown in Figure \ref{ngc22642}.

\begin{figure}
  \vspace*{174pt}
  \includegraphics{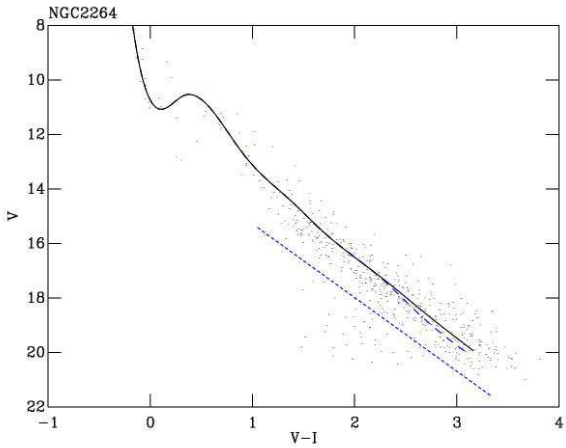}
  \caption{The effect of choosing a wider
    colour-magnitude selection in NGC2264. The effect is limited by the large
    number of sources within the sequence. The solid curve is the
    result using no selection, with the dotted line being the
    colour-magnitude selection and the dashed line the spline fit
    after the colour-magnitude selection is enforced i.e. only stars
    above the line were used to produce the spline curve.
    \label{ngc22643}}
\end{figure}

To remove BG and FG objects and spurious matches which lie below the
PMS a colour-magnitude selection was applied before fitting. We
removed objects lying below the dotted line in Figure \ref{ngc22643}.
We were concerned by the effect of this colour-magnitude selection on
the spline functions, so we examined fits using data with and without
a selection. The effect of removing the colour-magnitude selection for
NGC2264 is shown in Figure \ref{ngc22643}. It has little effect on the
resulting spline. 

The resulting fit is clearly an average through the spread in the
colour magnitude diagram \citep[sometimes referred to as an age
spread, see][and references therein]{2005MNRAS.363.1389B}.  
Part of this spread will be due to unresolved binaries which lie above
the single star sequence, lifting the spline above the single star
sequence.
Thus our splines do not necessarily follow the single-star sequence,
although the displacement due to binary should be the same for all
fields.
Furthermore the scale of this effect is small, as can be seen in the
CMDs were there is a clear separation between the binary and single
star sequence, e.g NGC2547 (Figure \ref{ngc2547}).

Sometimes the spline fit does not precisely follow the sequence the
eye picks out, for example $0.5<\textit{V-I}<1$ for NGC2264. Whether
the spline or the eye is correct turns out to be immaterial. The
deviations are never large enough to affect our conclusions.

\subsubsection{NGC2547}
\begin{figure}
  \vspace*{174pt}
  \includegraphics{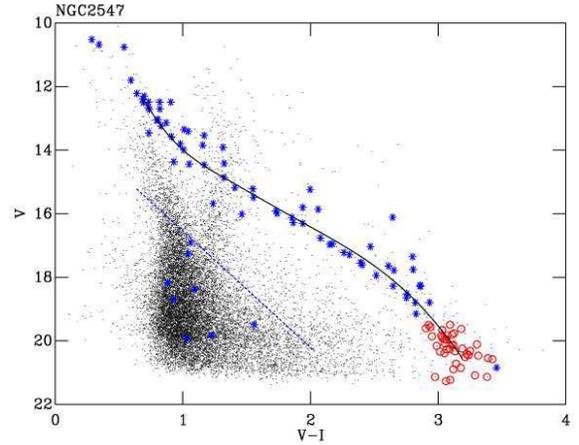}
  \caption{The full catalogue for the NGC2547 field (dots). 
    Asterisks are the X-ray sources from
    \citet{1998MNRAS.300..331J}, circles are spectroscopic members
    from \citet{2005MNRAS.358...13J}. The X-ray sources, some of which
    clearly lie in the BG MS contamination were removed using a
    colour-magnitude cut prior to fitting, shown here as the dotted
    blue line. \label{ngc2547}}
\end{figure}

The observations of NGC2547 were taken from
\cite{2002MNRAS.335..291N}, using Cousins \textit{I} ($\textit{I}_c$).
The photometry was calibrated with coefficients detailed in
\cite{2002MNRAS.335..291N} with the reddest standard having a
\textit{V-I} of 2.7. The resulting fit, after sequence selection is
shown in Figure \ref{ngc2547}. A colour-magnitude selection was used
to clip out probable non-members.  Only eight objects were removed,
all lying far from the sequence and within the contamination. The
fitted line lies slightly above the single star sequence, an effect
due to binary stars.

\subsubsection{ONC}
\begin{figure}
  \vspace*{174pt}
  \includegraphics{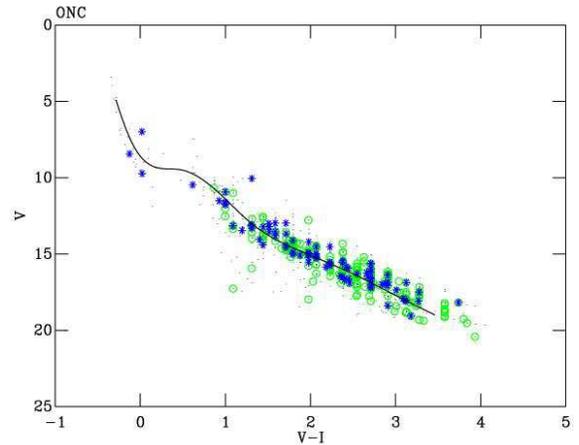}
  \caption{The ONC photometry from \citet{1997AJ....113.1733H}
  (dots). Circles are objects which are periodic variables from
  \citet{2002A&A...396..513H}. The asterisks are X-ray sources
  from \citet{2003ApJ...582..398F}. The best fit spline through all data
  points is also shown. \label{onc}}
\end{figure}

The photometry (using $\textit{I}_c$) and membership probabilities for
this region come from \cite{1997AJ....113.1733H}. For the ONC the
extinctions are calculated individually for each star, therefore the
order in which the fitting and conversion to absolute colours and
magnitudes is carried out is crucial. Where average extinctions have
been used the entire space is simply translated and the fitted spline
curve will not change shape. If however individual extinctions are
used shifts may therefore be different for each object, changing the
shape of the sequence. Consequently, for the ONC the extinctions must
be applied before the fitting procedure. This is also the case for the
later fields of IC348 and IC5146. The resulting sequence (extinction
corrected) with additional membership information fitted is shown as
Figure \ref{onc}.  All the stars shown were fitted to yield the spline
curve shown. No colour-magnitude selection was utilised here as nearly
all the stars within 15' of the centre are members of the ONC. This is
because the background stars are largely obscured by the dense cloud
cloud behind the cluster.

\subsubsection{NGC7160}
\begin{figure}
  \vspace*{174pt}
  \includegraphics{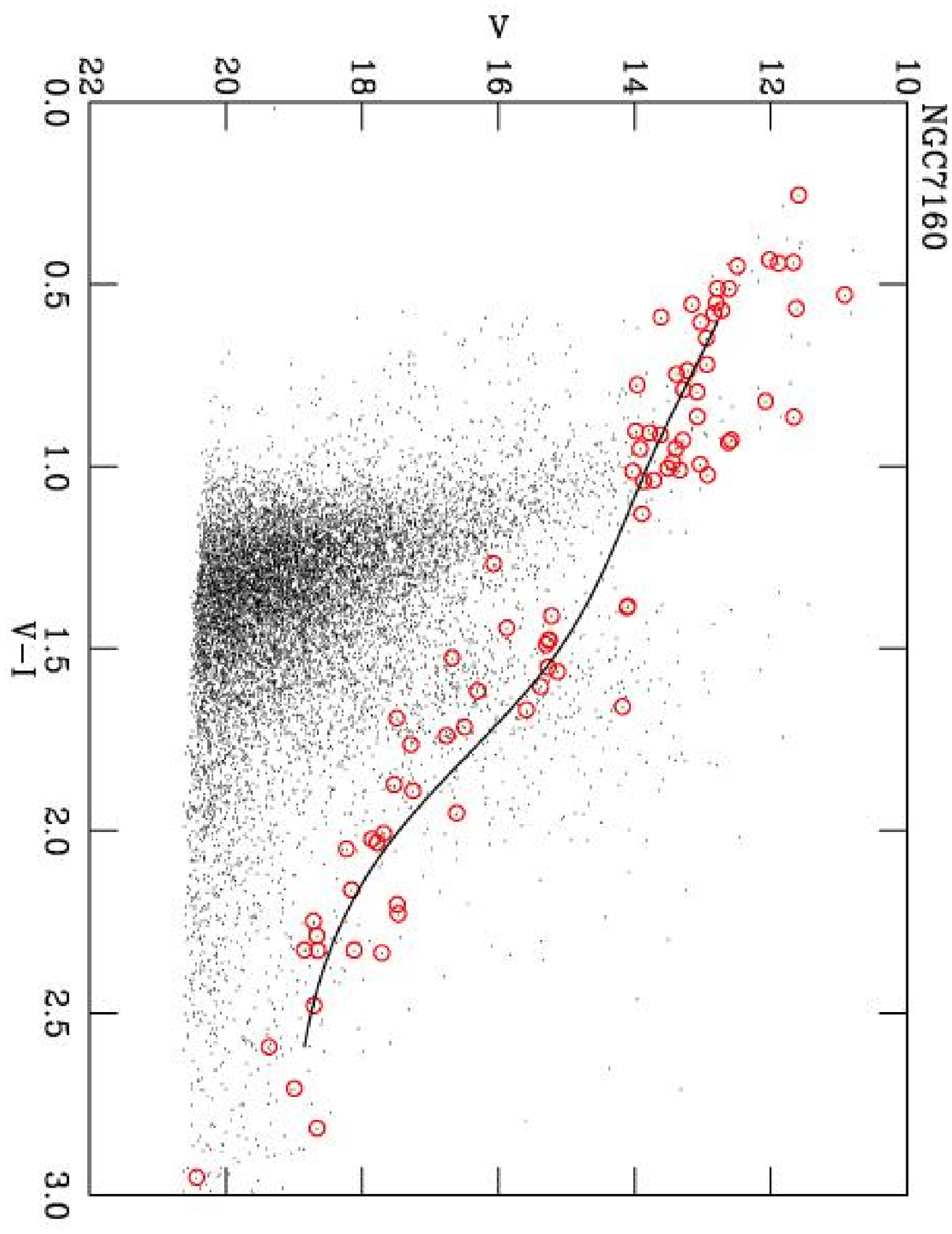}
  \caption{The full catalogue for the NGC7160 field (dots). 
    Circles are members from
    \citet{2004AJ....128..805S} and \citet{2005AJ....130..188S}.
    \label{ngc7160}}
\end{figure}

The sequence and fit is displayed in Figure \ref{ngc7160}.  The
members, although sparse, represent a relatively unbiased sample in CMD
space due to the `wide' criteria applied to assign memberships (see
Appendix \ref{ngc7160mem}).

\subsubsection{$\sigma$ Ori}
\begin{figure}
  \vspace*{174pt}
  \includegraphics{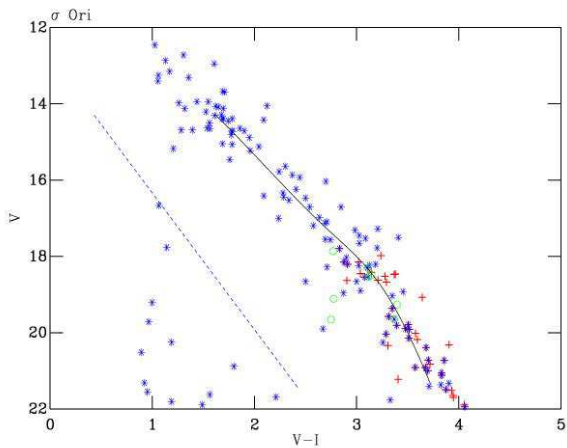}
  \caption{Stars selected as members in $\sigma$ Ori. 
    Circles are \citet{2005MNRAS.356.1583B}
    members. Asterisks are X-ray sources from
    \citet{2004A&A...421..715S}.  Crosses are members from
    \citet{2005MNRAS.356...89K}. The fitted spline curve is shown. The
    colour-magnitude selection is shown as the dotted line.
    \label{sori2}}
\end{figure}

For $\sigma$ Ori a colour-magnitude selection was used to clip out the
X-ray sources lying within the contamination. The resulting line fit
is displayed in Figure \ref{sori2}. The applied colour-magnitude
selection removed five objects all far from the sequence and within
the contamination.

\subsubsection{Cep OB3b}
\begin{figure}
  \vspace*{174pt}
  \includegraphics{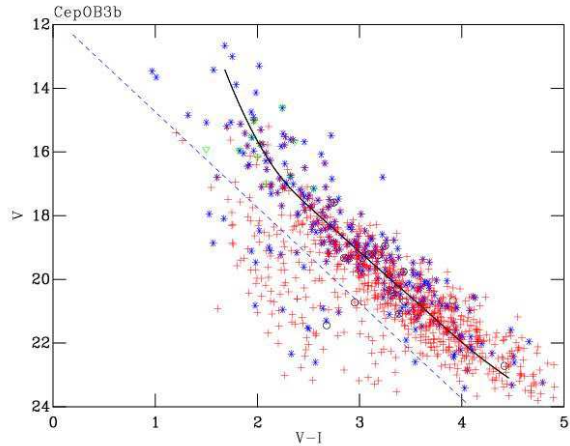}
  \caption{Stars selected as members in Cep OB3b. 
    Asterisks are X-ray sources from \citet{1999MNRAS.302..714N}, 
    \citet{2006ApJS..163..306G} and the second PSPC catalogue.
    Triangles are members from
    \citet{2003MNRAS.341..805P}. Circles are H$\alpha$ sources from
    \citet{2002AJ....123.2597O}. Crosses are the periodic variables
    from Littlefair et al (in preparation). The fitted spline curve is
    shown.
    \label{cepOB3b2}}
\end{figure}

The selected members and fit for Cep OB3b are shown as Figure
\ref{cepOB3b2}. A colour-magnitude selection has again been applied,
although as in Section \ref{fitting}, since most objects lie in the
sequence the effect on the fit is negligible.

\subsubsection{IC348}
\begin{figure}
  \vspace*{174pt}
  \includegraphics{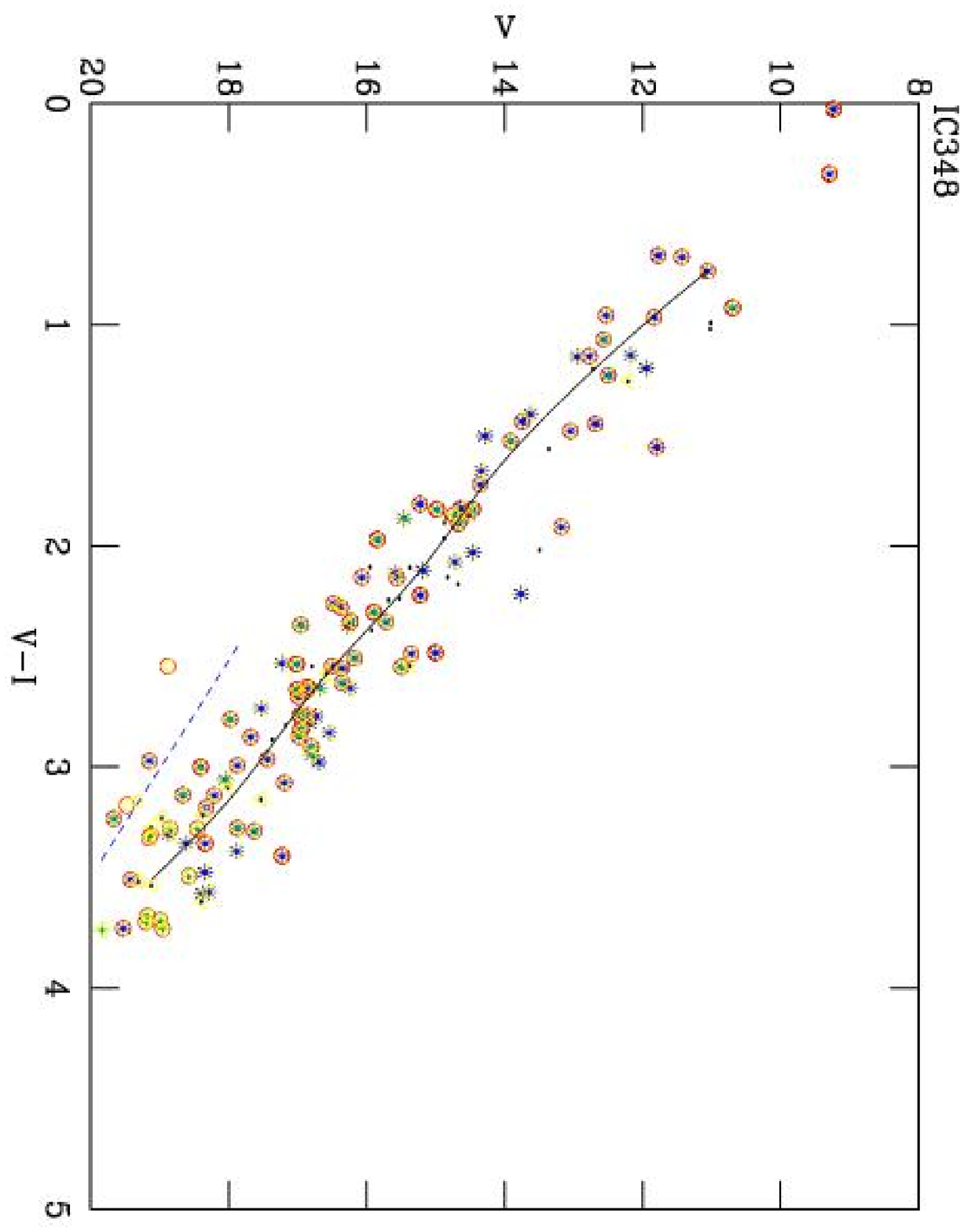}
  \caption{Stars selected as members in IC348. The asterisks are X-ray sources from
    \citet{2002AJ....123.1613P} and the Second ROSAT PSPC catalogue.
    Circles are the periodic variables from
    \citet{2004AJ....127.1602C} and \citet{2005MNRAS.358..341L}.
    Crosses are H$\alpha$ sources from \citet{1998ApJ...497..736H}.
    Triangles are spectroscopic members with extinctions from
    \citet{2003ApJ...593.1093L} and \citet{1998ApJ...497..736H}.
    Individual extinctions from \citet{2003ApJ...593.1093L} and
    \citet{1998ApJ...497..736H} have been applied before fitting.
    \label{ic3482}}
\end{figure}

For IC348 individual extinctions were available for all but 21 of the
members. Therefore only those members with extinctions have been used.
A colour-magnitude selection has been applied removing five stars below
the sequence. The individual dereddenings were applied prior to
fitting. The dereddened sequence with the fitted curve can be seen as
Figure \ref{ic3482}.

\subsubsection{$\lambda$ Ori, NGC2362, IC5146 and NGC6530}
\begin{figure}
  \vspace*{174pt}
  \includegraphics{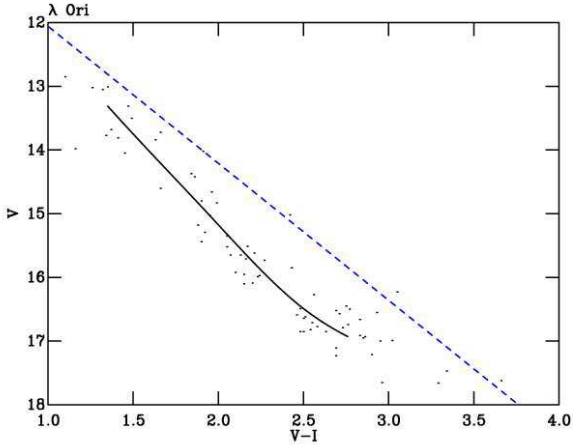}
  \caption{Stars selected as members in $\lambda$ Ori. The dots are Li members from
    \citet{2001AJ....121.2124D}. \label{lambdaori}}
\end{figure}

\begin{figure}
  \vspace*{174pt} 
 \includegraphics{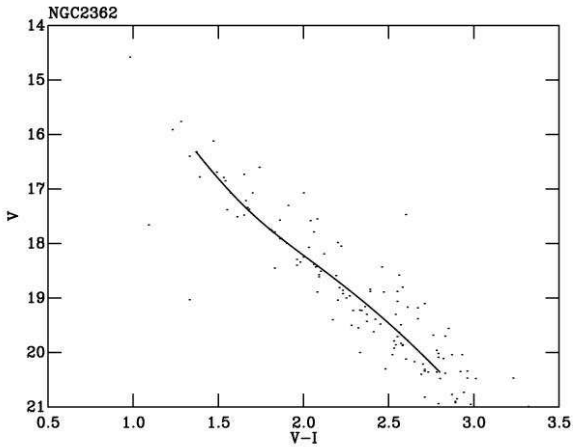}
  \caption{Stars selected as members in NGC2362. 
    The dots are H$\alpha$ and spectroscopically confirmed PMS members from
    \citet{2005AJ....130.1805D}. \label{ngc2362}}
\end{figure}

\begin{figure}
  \vspace*{174pt}
  \includegraphics{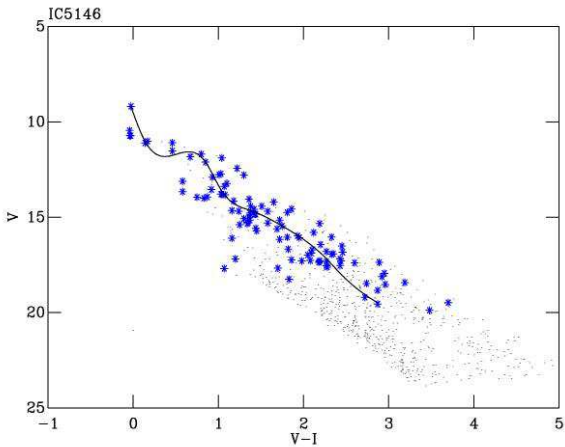}
  \caption{Stars selected as members in IC5146. 
    Dots are likely PMS members from \citet{2002AJ....123..304H}. 
    Asterisks are H$\alpha$ members from \citet{2002AJ....123..304H}
    with known spectral types excluding stars lying below the Pleiades MS. 
    Only the asterisks were fitted. \label{ic5146}}
\end{figure}

\begin{figure}
  \vspace*{174pt}
  \includegraphics{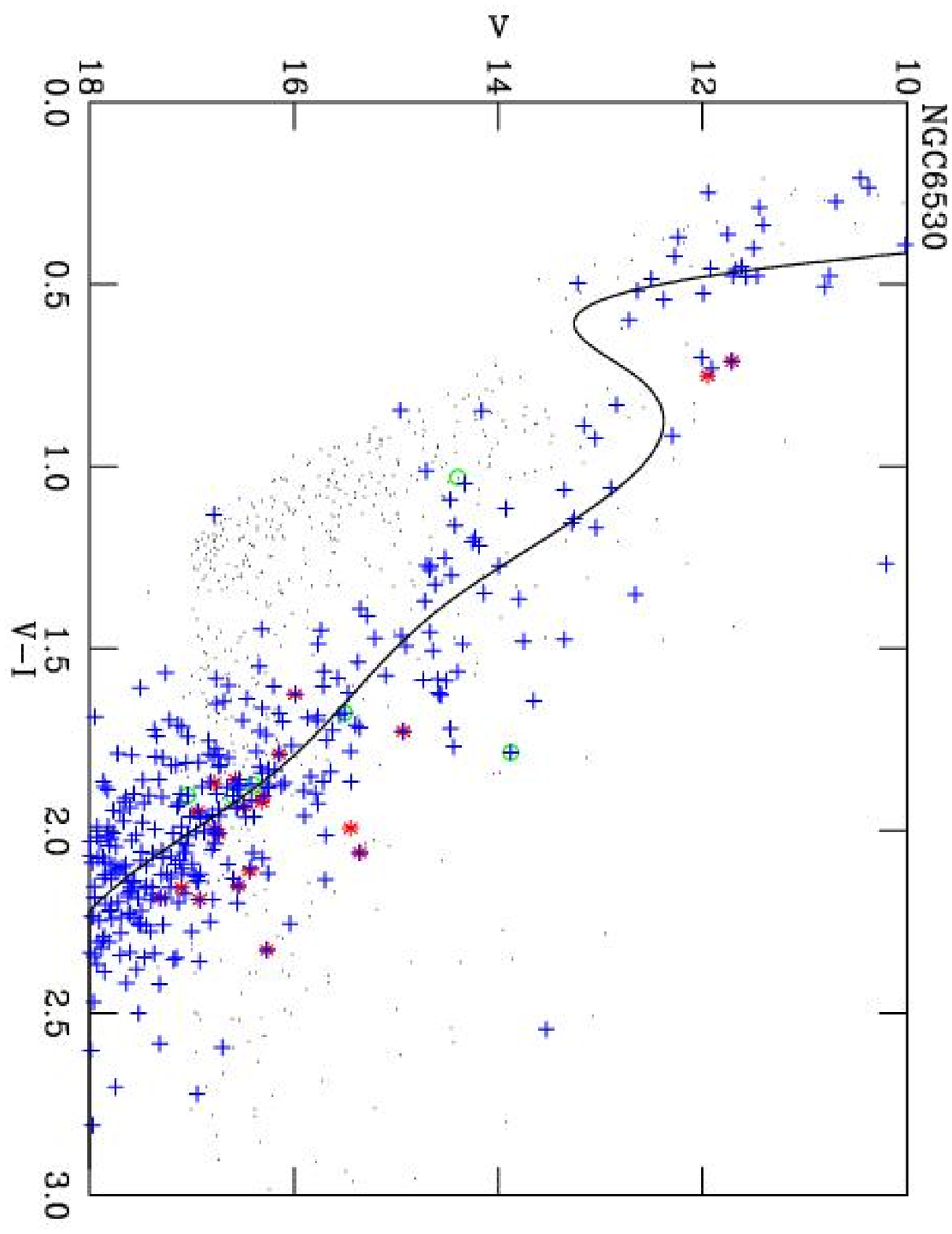}
  \caption{The photometry of NGC6530 from \citet{2005A&A...430..941P} 
  (dots). 
  Crosses are X-ray sources from \citet{2005A&A...430..941P}. 
  Circles are weak H$\alpha$ stars from \citet{2000AJ....120..333S}. 
  Asterisks are strong H$\alpha$ stars from
  \citet{2000AJ....120..333S}. 
  \label{ngc6530}} 
\end{figure}

The data for $\lambda$ Ori are from \cite{2001AJ....121.2124D} taken
in the Johnson-Kron-Cousins system and calibrated to Landolt
standards, a colour-magnitude selection has been applied removing
three stars lying well above the sequence. The data for NGC2362 are
from \cite{2005AJ....130.1805D}, taken with an \textit{$I_c$} filter
and tied to Landolt standards. The Cousins photometry for IC5146 comes
from \cite{2002AJ....123..304H}, here only the stars shown as
asterisks  in Figure \ref{ic5146} were fitted.
This contains photometry from two areas approximately 10' apart. The
photometry for NGC6530 is from \cite{2005A&A...430..941P}.  The
members and spline fits for these fields can be seen as Figures
\ref{lambdaori}, \ref{ngc2362}, \ref{ic5146} and \ref{ngc6530}.

\subsection{Isolating the sequence for h and $\chi$ Per}
\label{Positional_isolation}
In this field there was no membership data of the form used above, so
colour-magnitude selections were employed. In addition as the group in
question is clustered in the FOV it was possible to remove a
proportion of FG and BG objects by selection of a limited area of the
FOV. We however needed a guiding mechanism to allow us to pick the
approximate central coordinates and a subsequent area to isolate.  The
following procedure was undertaken to isolate the cluster in celestial
coordinates.

\begin{figure}
  \vspace*{174pt}
  \includegraphics{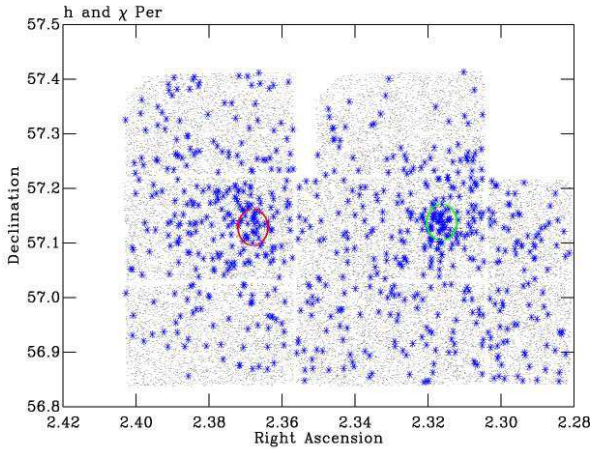}
  \caption{The positions of stars in the h and $\chi$ Per catalogue.  
  Asterisks are stars with $\textit{V}<14$th, and the large
  circles show the regions we selected.} 
\label{persei2}
\end{figure}

\begin{figure}
  \vspace*{174pt}
  \includegraphics{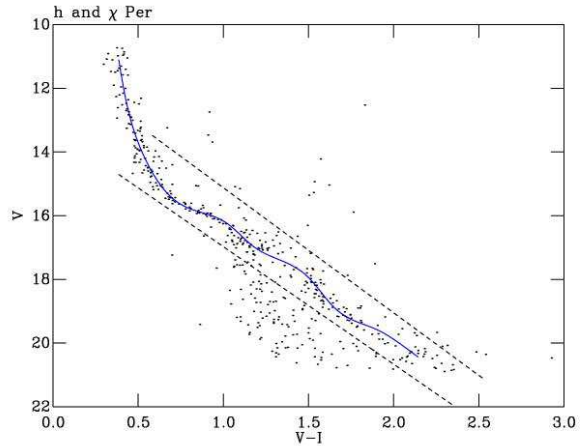}
  \caption{Stars within 2 arcmin radius of the centres of h and $\chi$
    Per. 
    Stars in the h Per region have been shifted in extinction to match
    those  $\chi$ Per. 
    The dotted lines are the colour-magnitude cuts used, and the solid
    curve the best fitting spline.
    \label{persei3}}
\end{figure}

\begin{enumerate}
\item \textbf{Lead Stars.} We show our CMD for the full h and $\chi$ Per
  catalogue in Figure \ref{persei1}. The sequence is clearly visible
  and the brighter stars clearly lie blue-ward of the contamination.
  Therefore the positions of these stars on the sky were plotted,
  allowing us to trace the outline of the clusters. This can be seen
  in Figure \ref{persei2}.
  
  Three other sets of lead star candidates were explored.
  \cite{2002PASP..114..233U}; proper motion members or photometric
  members from \cite{2002ApJ...576..880S}; and
  \cite{1996yCat.5027....0M}. The coordinates from each set were
  carried forward through the following steps. However the central
  coordinates for all cases were very similar.

\begin{figure}
  \vspace*{174pt}
  \includegraphics{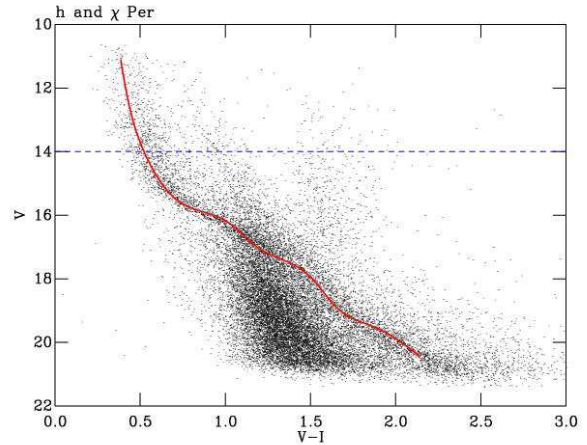}
  \caption{The full catalogue for the h and $\chi$ Per field (dots). 
  The dotted line shows a colour-magnitude selection of $\textit{V}<14$th, 
  showing the sequence clear of the contamination. The solid line is
  the best fitting spline to the selected sequence corrected to the
  mean extinction for the two clusters.
  \label{persei1}}
\end{figure}

\item \textbf{Positional selection.}  A circular positional selection was
  applied to the full catalogue around the central coordinates, with
  an inclusion radius varying from 5' down to 1' in increments of 1'.
  The CMDs for each were examined and a `by eye' selection made for
  the CMD yielding the clearest sequence, optimising the member to
  field-object ratio. The best results were obtained with a radius of
  2'.  The final central coordinates for each cluster are: $\alpha$=2
  22' 5.02'' $\delta$=+57 7' 43.44'' ($\chi$ Per), $\alpha$=2 18'
  58.76'' $\delta$=+57 8' 16.54''(h Per), J2000.
  
Since the measured ages in virtually all the literature for h and
$\chi$ Per are the same, it was possible to increase our confidence
of selecting the correct sequence in CMD space by combining the CMDs
from both clusters. To achieve this any differences in extinction
and distance modulus must be accounted for. We derived extinctions
and distance modulus shifts in the sense h Per minus $\chi$ Per
using three methods.
\hfil\break 
(a) Matching the curve in the MS at \textit{V}=16 mags for both
clusters by eye, we found $A_V=0.08\pm0.05$, $d_m=0.07\pm0.05$ mags.
\hfil\break 
(b) Matching the centre of the UMS (Upper Main Sequence) of $\chi$
Per to the centre of the more scattered UMS in h Per gave a shift in $|d_m|$
of $<0.05$, with a shift in extinction of $A_V=0.2\pm0.05$.
\hfil\break 
(c) The Q-method (see Section \ref{Q}) gave $A_V=0.157\pm0.0212$.
\hfil\break 
We chose the Q-method value for the extinction and no shift in $d_m$ .
The scatter in the different derivations suggests an uncertainty of
approximately $\pm0.05$ in extinction and distance modulus.  This
shift was applied to h Per, and the combined catalogue used from
thereon (Table 28).  The final area selected can be seen overlaid on
Figure \ref{persei2}.

\item \textbf{Colour-magnitude selection and fitting.}  The final
  sequence was then fitted as described in Section \ref{fitting}. Cuts
  in colour-magnitude space were then used to clip out stars lying far
  from the sequence.  The cuts were varied until a best fit of the
  spline curve to the sequence was achieved.  The final sequence, cuts
  and spline curve are shown in Figure \ref{persei3}. To further
  increase our confidence in selection of the correct sequence the
  best fit spline was shifted to the mean extinction for both clusters
  and overlaid on the CMD showing the full catalogue. The spline curve
  clearly follows the sequence one would select by eye from this CMD,
  see Figure \ref{persei1}.
\end{enumerate}

\section[]{Parameters}
\label{Parameters}
To compare the data for our sequences on a single CMD we required
estimates for the extinction ($A_V$) and true distance modulus
($d_m$). To obtain a self-consistent set of empirical isochrones it is
clearly better to use a single method to determine extinction. This
was possible for a range of fields using the Q-method as detailed
below. Where data for the Q-method were unavailable we have used
values from the literature, given in Appendix \ref{appendix2}. A
summary of the distance moduli and extinctions we used can be found in
Table \ref{dm_AV}.

\subsection{Q-method}
\label{Q}
Using \textit{UBV} photometry from the literature we applied the
Q-method of \cite{1953ApJ...117..313J}. This method uses the colour
relationships of OB stars and is only valid over a small range of
spectral types. \cite{1953ApJ...117..313J} quote
(\textit{B-V})\textit{$_0$}$=-0.009+0.337Q$ to be valid for
$-0.80<Q<-0.05$, converting to an intrinsic colour range of
$-0.28<$(\textit{B-V})\textit{$_0$}$<-0.03$. This was used in
conjunction with the initial parameter estimates to calculate an
apparent colour range over which the method was applicable in each
field.  We then derived a mean extinction which we could in turn apply
to our selection. The loop of applying a mean extinction, calculating
$Q$ from the stars within the colour range and calculating a new mean
extinction was iterated until the number of stars used in successive
iterations remained constant. The RMS around the mean value divided by
the square root of the number of points provides an uncertainty.

\subsubsection{h and $\chi$ Per}
\begin{figure}
  \vspace*{174pt}
  \includegraphics{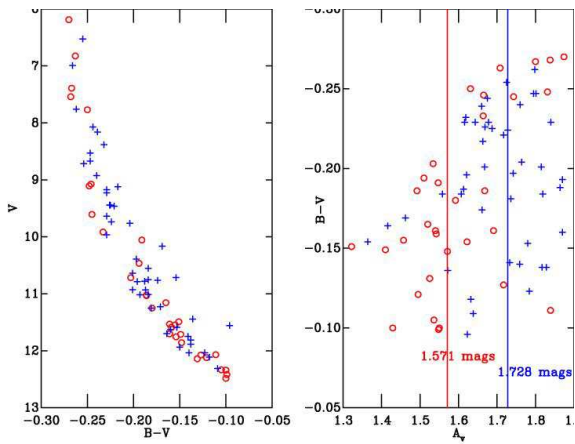} 
  \caption{\textit{Left}. The sequences for h and $\chi$ Per O and B stars
  after the application of the Q-method. \textit{Right}. The extinction. The
  circles are $\chi$ Per and the crosses are h Per. The resulting
  median values of $A_V$=1.571 and 1.728 for $\chi$ and h Per
  respectively are shown. \label{doubleq}}
\end{figure}

Here \textit{UBV} photometry was obtained from
\cite{2002ApJ...576..880S}. In Section \ref{Positional_isolation} we
limited our study to two circular fields centred on the clusters. As
extinction varies across the field of view for these two clusters, we
have limited ourselves to stars from \cite{2002ApJ...576..880S} lying
within these areas. The result is shown as Figure \ref{doubleq}. The
difference of the median values of extinction using Q between the two
positionally selected areas was found. The shift derived for h Per was
$A_V=0.157$ which has already been applied in Section
\ref{Positional_isolation}. Therefore the reddening to $\chi$ Per only
was needed, being $A_V=1.571 \pm0.079$.

\subsubsection{NGC2547, NG2264 and Cep OB3b}
\label{q_ext}
For NGC2547 we used the photometry of \cite{1982A&AS...47..323C}. The
resulting median extinction value was $0.157\pm0.008$ mags. For
NGC2264 \textit{UBV} member photometry from \cite{1980MNRAS.190..623M}
yielded a median $A_V=0.371\pm0.0409$. The final field is Cep OB3b,
where the photometry was from \cite{1996A&A...312..499J}, with a
resulting median value of $A_V=2.88\pm0.039$.

\section{Data comparison}
\label{Data_comparison}
These data present an opportunity to make a qualitative assessment of
how the PMS changes with age. To enable us to do this the sequences
were shifted by the adopted distance moduli and extinctions (see
Figure \ref{gap_four}). We selected four of our best sequences
spanning as large an age range as possible. Data for NGC2264 has been
supplemented by photometry from \cite{1980MNRAS.190..623M}. They used
radial velocity to assign membership probabilities, and we plot only
stars with membership probabilities of $>90\%$. This allowed us to
plot the sequence members above \textit{$V_0$}=11, where the stars are
saturated in our catalogue.

\subsection{Sequence spreads}
The sequences in Figure \ref{gap_four} appear to ``crystallise'' with
age. The absolute spread in CMD space of the sequences reduces as the
age of the field increases. The trend appears to continue in the
fields older than those presented in Figure \ref{gap_four}, with
NGC2547 ($\simeq$30 Myrs) showing a particularly clear sequence.
Sequences older than ~5 Myrs do not appear to show large spreads.
However, sequences younger than this show large spreads not explicable
by photometric errors, variability and binarity, as shown in the case
of Cep OB3b and $\sigma$ Ori by \cite{2005MNRAS.363.1389B}. This
contrast with the Upper Sco-Cen OB association at $\simeq5-6$ Myrs,
which has an apparently wholly explicable photometric spread
\citep{2002AJ....124..404P}.  It appears there is a change at about 5
Myrs.

This suggests a possible cause of the anomalous spreads; namely
accretion or infall history. The half-life for accretion discs is
estimated to be 3 Myrs, therefore for low-mass objects one would
expect the discs to have disappeared by $\simeq6$ Myrs
\citep{2006MNRAS.369..272O, 2001ApJ...553L.153H}. Accretion, from a
disc increases the continuum luminosity along with various other
effects, possibly pushing the stars blue-ward, resulting in a scatter
both temporally older and younger \citep{1999MNRAS.310..360T}.  This
component of the spread would not be present in sequences older than
$\simeq6$ Myrs. It should be noted however that this half-life is
derived from the presence of dust; the half-life of gas within these
discs is less certain.    

An obvious alternative is that the groups are not co-eval. The
timescale of star formation itself must produce an intrinsic spread in
age and hence in position in the CMD.  In addition there is strong
evidence for the episodic star formation within a group, sub-group,
cluster or association. Thus another explanation for the secular
evolution of the sequence spreads we observe is in Figure
\ref{gap_four} could be an initial age spread evolving to older ages
and the absolute spread therefore reducing as the isochrones bunch up
in CMD space, i.e.  $d(age)/dV$ increases. If the spread within the
CMD is due to a genuine age spread, since our empirical isochrones are
based on median magnitudes they will be biased toward younger ages.

\subsection{Evidence for a gap in a \textit{V}, \textit{V-I} CMD}

\begin{figure*}
  \vspace*{348pt}
  \includegraphics{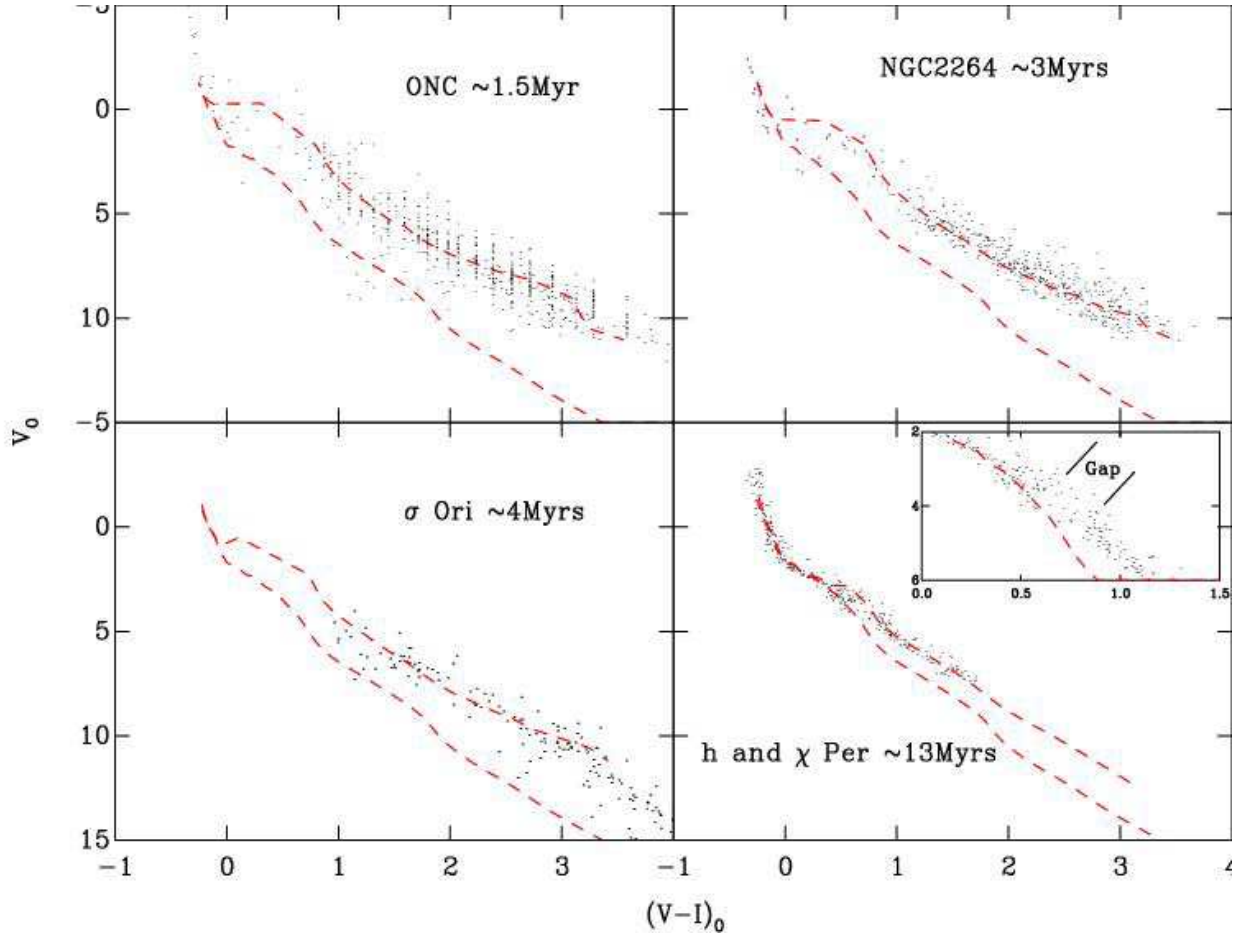}
  \caption{The likely PMS objects for fields of different ages, 
    plotted in dereddened colours and absolute magnitudes. The spread
    in absolute colour magnitude space clearly decreases as age
    increases. The ZAMS and isochrones of approximately the correct age from
    \citet{2000A&A...358..593S} are shown.}
\label{gap_four}
\end{figure*}

The ONC has a discernible gap between the PMS stars and the MS stars,
at the point where the PMS isochrone meets the zero age main sequences
(ZAMS). This can be seen clearly in Figure \ref{gap_four}. There also
appears to be a gap, again where the PMS meets the MS in h and $\chi$
Per and NGC2264 (Figure \ref{gap_four}).  A gap is evident in the
photometry for NGC6530 and perhaps IC5146 (see Figures \ref{ngc6530}
and \ref{ic5146}).  A further excellent example of the gap can be seen
in the CMD of NGC4755 in \cite{2006astro.ph..3009L}.

To ascertain whether this gap between MS and PMS stars is a general
feature we used the isochrones to find the position (in CMD space) of
the PMS-MS connection.  In some of our younger fields the gap is
situated around our bright magnitude limit. In these cases we cannot
be sure of the gap's existence. There is however a discernible dearth
of members at the head of the PMS, which can be seen in Cep OB3b,
$\sigma$ Ori, IC348, NGC2362 and $\lambda$ Ori. This explains the
general shape of many CMDs of PMS regions, which appear to consist of
a sequence parallel and redward of the contamination which disappears
at brighter magnitudes.  The brightest stars at the head of this
sequence must represent the top of the PMS, with the MS appearing
blue-ward of this.  The MS, and therefore the gap, is often obscured by
contamination.

We could not definitively detect the gap in NGC7160 or NGC2547. In the
case of NGC7160 the gap would fall in an area devoid of membership
information. Finally for our oldest field NGC2547 the PMS and MS
isochrones lie very close together meaning the gap, if present, would
be very small compared to the younger fields.

It is apparent that this phenomena occurs in all datasets where we
would expect to observe it.  This feature delineates the transition
from PMS to MS, hence \cite{2004AJ....128..765S} call this the PMS/MS
transition region (they identified it in a \textit{$J_s$},
\textit{$J_s-K_s$} CMD of NGC3603).  We prefer the term radiative
convective (R-C) gap as we feel this emphasizes the physics involved
which we outline in the next section.

\subsection{The origin of R-C gap}
\begin{figure}
   \vspace*{174pt}
  \includegraphics{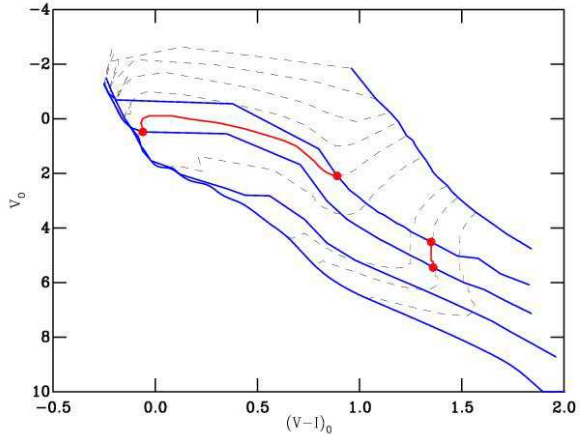}
  \caption{
    Isochrones (solid blue lines) from \citet{2000A&A...358..593S} for
    the approximate ages of the four fields in Figure \ref{gap_four};
    1, 3 and 13 Myrs.
    Mass tracks (dashed lines) are show for  7, 6, 5, 4, 3, 2, 1.2,
    1.0, and 0.8$M_{\sun}$.
    The evolution between 1 and 3 Myrs of 3 and 1$M_{\sun}$ stars are
    shown as filled circles, with the track highlighted in red. 
    \label{theory}}
\end{figure}

From the theoretical isochrones an explanation for this gap becomes
apparent (see Figure \ref{theory}).  Physically the R-C gap is the
transition phase of stars from a fully convective object to the
development of a radiative core. The Hayashi tracks which the lower
mass PMS stars descend in CMD space represent the gravitational
contraction of a fully convective object.  As the star heats and the
core density gradually increases, at some stage the core becomes
radiative, and the track changes to a Henyey track which is almost
horizontal in a \textit{V}/\textit{V-I} CMD. For massive stars this
happens earlier in the star's evolution than for low-mass stars
\citep{1989fsa..book.....C}; a 7$M_{\sun}$ star will develop a
radiative core almost immediately.

The development and evolution of the radiative core causes the
separation of the PMS and MS. As the transition proceeds the mass
tracks are almost parallel with the isochrones, meaning evolution
across the area of CMD space between the PMS and MS is rapid compared
to the descent down the Hayashi tracks. The difference in motion in
CMD space is clearly demonstrated in Figure \ref{theory}.  Here the
distance between a 1 Myr and 3 Myr isochrone for a star on a fully
convective track is small when compared to a star at the head of the
PMS which will have joined the MS by 3 Myrs, moving much farther in
CMD space. This means at any given age a sequence should have a
relative sparsity of stars in this gap compared to the slowly evolving
PMS and almost stationary MS (on these timescales).

As is visible from the data in Figure \ref{gap_four} and the theory in
Figure \ref{theory} the size of the dislocation between the MS and PMS
is a function of age. For the younger fields the PMS lies far from the
MS and the transition occurs at high masses. The higher the mass of a
star the earlier the formation of a radiative core and the sooner it
changes to a Henyey track. Higher mass objects move to the MS more
swiftly along these tracks. The reverse is also true. As the field
becomes older, the mass at which the core develops falls and the
evolution of the star progresses more slowly.

The age dependency of the size of the gap has important ramifications.
If it is possible to identify the head of the PMS and the MS members
for a given field, the separation of these will be a distance and
reddening independent measure of the age. There are several
restrictions. If the field is older than around 15 Myrs the gap
becomes too small to detect reliably. The gap may also lie within an
area of high contamination, meaning many objects will need to be
assessed for membership to be certain of detecting the gap. In some
cases variable extinction will prevent photometric detection of the
gap, as the relative positions of the stars shift in a CMD; in these
cases individual extinctions will need to be found. As the size of the
gap depends critically on the formation and growth of the radiative
core, it may also represent an excellent test of the stellar interior models.

\section{Relative ages}
\label{Relative_ages}
\begin{figure}
  \vspace*{174pt}
  \includegraphics{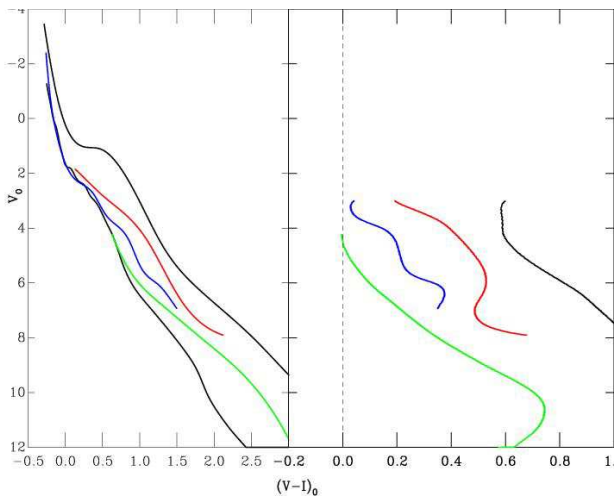}
  \caption{A sample of the empirical isochrones. Youngest to oldest, the
    ONC, NGC7160, h and $\chi$ Per and NGC2547. The \textit{left}
    panel is the empirical isochrones in an absolute \textit{V},
    \textit{V-I} CMD, with the lowest curve the ZAMS relation of
    \citet{2000A&A...358..593S}. The \textit{right} panel is the ZAMS
    subtracted space; at a given \textit{V} the ZAMS \textit{V-I} is
    subtracted from the \textit{V-I} of the spline. The vertical
    dashed line shows the position of the ZAMS}
\label{spline_sample}
\end{figure}

The sequences displayed in Figure \ref{gap_four} are the results after
membership selection.  To create a relative age ladder however the
sequences must appear on the same CMD, superimposed over one another.
Such a figure would be confused due to the number of points and the
spreads of the sequences. Therefore at this stage we move to
representing the sequences by the spline fits. To aid analysis we also
created a ZAMS subtracted CMD, using the ZAMS relation of
\cite{2000A&A...358..593S}.  In these diagrams the points on the
spline fits each have the \textit{V-I} of the MS star at equivalent
\textit{V} subtracted from their \textit{V-I}. This separates the
sequences much more clearly. Figure \ref{spline_sample} shows a sample
of the empirical isochrones.

The disadvantage of the ZAMS subtraction process is that it enhances
the residuals of the spline fitting procedure.  As can be seen in
Figure \ref{tail_four} at the R-C gap there is a dearth of data which
causes the spline fit to degrade. This also causes the fit to the MS
after the R-C gap to be poor as the spline fits enforce a gradient
continuity criterion. These residuals have been acceptable (if noted)
until now but are greatly magnified in the ZAMS subtracted space.
Therefore we have cut the splines in ZAMS subtracted space at a
\textit{$V_0$}$=3$.

\subsection{Theory comparison}
It is useful at this stage to compare the empirical isochrones we have
created to their theoretical counterparts. Although we have used
isochrones from \cite{2000A&A...358..593S}, the effects described are
apparent with all the other isochrones we have examined, though a
complete investigation of all the models lies outside the scope of
this paper.

We have adopted an approximately solar metallicity, $Z=0.02$.
\cite{2006A&A...446..971J} shows that for a a number of star forming
regions the metallicty is slightly sub-solar, it is also likely that
different clusters, associations, groups or sub-groups will have
differing compositions.  Therefore it is pertinent to ask what age
difference would be obtained by fitting a higher or lower metallicity
isochrone to our representative sequence of 3 Myrs. If we compare a 3
Myr $Z=0.02$ isochrone with isochrones of $Z=0.01$ and $Z=0.04$ the
closest matches have approximate ages of 2.2 and 3.5 Myrs
respectively.  Thus changes in the metallicity by a factor two
are too small to affect the discussion which follows.

In Figure \ref{spline_sample} we can see a minor problem. NGC2547 can
be seen to cut across the younger sequences, this is probably due to
observations being made using a different \textit{I} filter (see
Section \ref{literature_data}) coupled with the fact that the reddest
standard is at a \textit{V-I} of 2.5 in this case, exaggerated by the
ZAMS subtraction process.

\begin{figure*}
  \vspace*{348pt}
  \includegraphics{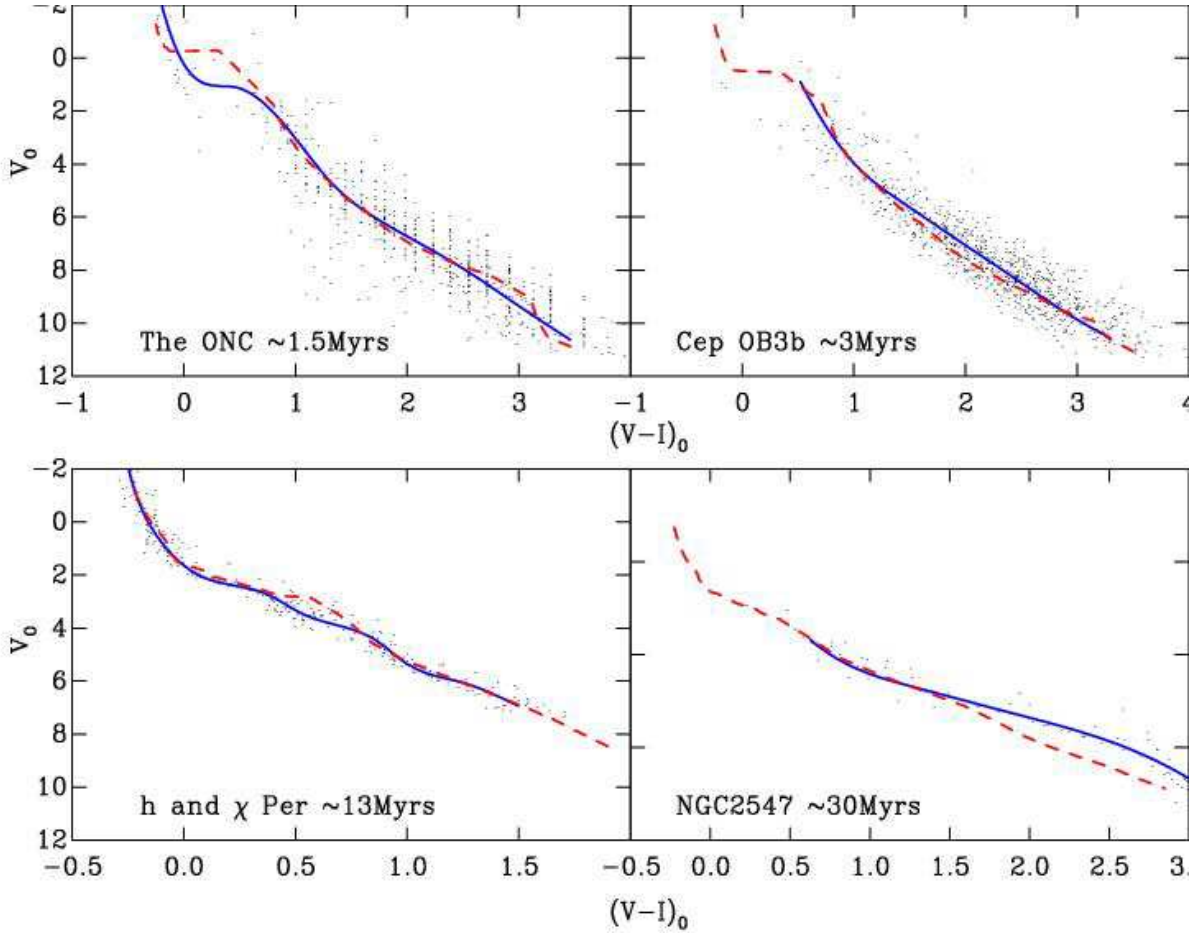}
  \caption{The selected members for the ONC, Cep OB3b, h and $\chi$
    Per and NGC2547. 
    The solid line is the empirical isochrone. 
    The dashed line is the best fit isochrone (by eye)
    from \citet{2000A&A...358..593S} ($\simeq1.5$, $\simeq3$, $\simeq13$
    and $\simeq16$ Myrs respectively). The sequences and empirical
    isochrones can be seen to move away from the theoretical
    counterparts to varying degrees. \label{tail_four}}
\end{figure*}

In addition, our comparison reveals three serious concerns.  (i)
Figures \ref{gap_four} and \ref{tail_four} show that whilst an
acceptable fit to the theoretical isochrones can be obtained over most
of the sequence, the model isochrone lies above the data at the
boundaries of the R-C gap for NGC2264 and h and $\chi$ Per.
Furthermore the MS members overlap the PMS in magnitude space for all
our fields, in direct contradiction to the isochrones, assuming a
coeval population (see Section \ref{conclusion} for a further
discussion of this). (ii) For younger sequences a theoretical
isochrone fitted to the PMS is a poor match to the MS.  (iii) Although
the young sequences have a large scatter, the PMS isochrones fail to
consistently follow the centre of this spread.  For example, in the
NGC2264 most stars lie above the isochrone at \textit{V-I}$>1$, but
below it at \textit{V-I}$<1$.  The conclusion is that when fitting
isochrones to sequences across a limited colour range, one must be
aware that the fit will not necessarily be good across the rest of the
sequence.  This implies that the age derived depends on which part of
the sequence is fitted.  This is why empirical isochrones are
potentially superior to their theoretical counterparts.

\subsection{Selecting the fiducial sequences}
\label{fids}
As we shall see later, we can roughly group the youngest objects into
three groups with ages of 1, 3 and 5 Myrs. The sequences older than
this are much more clearly separated. These three groups are separated
in age such that using the maximum uncertainty in distance modulus
(typically 0.2 mags) could shift a sequence by one group, older or
younger.  For use as fiducial sequences we select the spline curves of
one sequence from each group, typifying the group. We preferentially
select the sequence from each group with the most certain distance and
age. These fiducial isochrones can then be used to derive relative
ages to further fields, with selection of the fiducials guided by the
sequence's placement in colour-magnitude space. The group at around 5
Myrs is problematic, as discussed later it contains only two clusters
$\sigma$ Ori and IC348; $\sigma$ Ori has an uncertain distance.
However, IC348 is found to be much older (in this work) than the age
given in the literature. Therefore $\sigma$ Ori has been chosen as the
fiducial for this group, despite its large distance uncertainty.

\begin{figure}
  \vspace*{174pt}
  \includegraphics{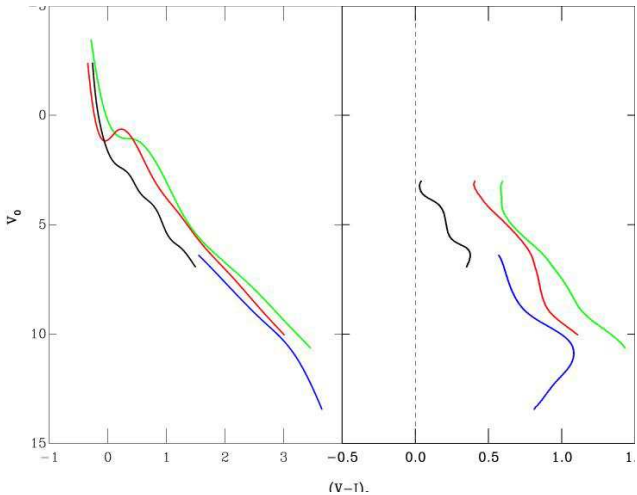}
  \caption{The selected fiducial sequences; h and $\chi$ Per, $\sigma$
    Ori, NGC2264 and the ONC. The vertical dashed line shows the
    position of the ZAMS}
\label{fiducials}
\end{figure}

The fiducial empirical isochrones selected were the ONC, NGC2264,
$\sigma$ Ori and h and $\chi$ Per, with assumed ages (from the literature or
fitting of \cite{2000A&A...358..593S} isochrones) of 1, 3, 4 and
13 Myrs. 
These fiducials will be used in Section \ref{ladder_use}. 
The fiducials are shown in Figure \ref{fiducials}.

\section{Using a relative age ladder}
\label{ladder_use}

We can now use our age ladder to determine relative ages for several
fields.  Throughout the figures in this section the fiducial
isochrones are marked as dashed lines, with the subject sequence
appearing as a solid line.

\subsection{Cep OB3b}
\label{cepOB3b_discuss}
For Cep OB3b the membership list has been discussed in Section
\ref{CepOB3b_members} and the CMD with the members overlaid is shown
as Figure \ref{cepOB3b1}. Estimates for the distance modulus are from
Section \ref{cepOB3b_dist} and the extinction from Section
\ref{q_ext}.

\begin{figure}
  \vspace*{174pt}
  \includegraphics{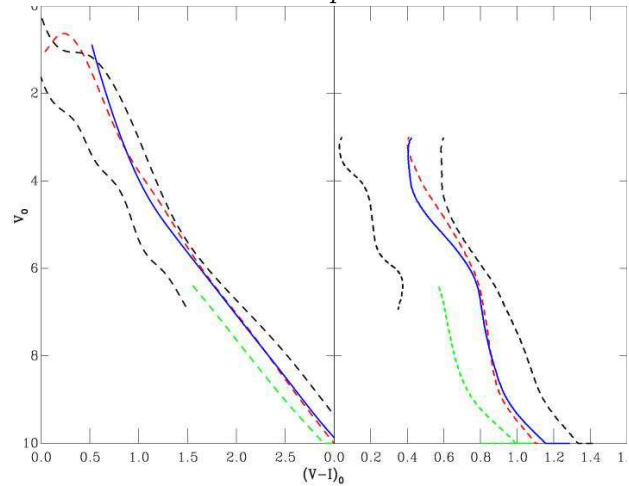}
  \caption{The fitted sequence for Cep OB3b (solid line). 
    NGC2264, h and $\chi$ Per,
    $\sigma$ Ori, and the ONC are shown as dashed lines for reference.
    \label{cepOB3bage}}
\end{figure}

\begin{figure}
  \vspace*{174pt} 
  \includegraphics{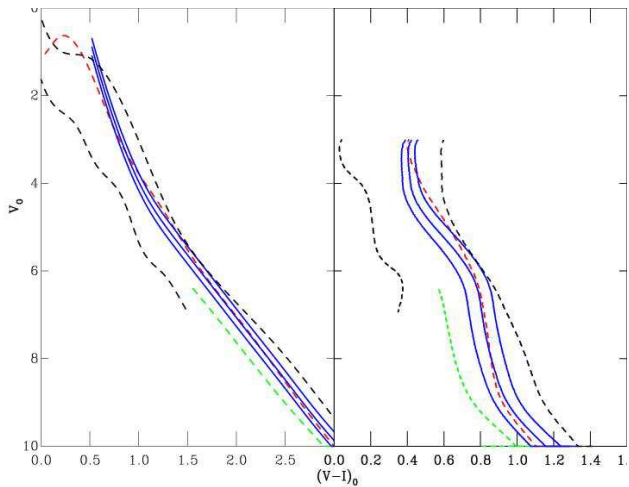}
  \caption{The sequence for Cep OB3b assuming thee different
    distance moduli (solid lines). NGC2264, h and $\chi$ Per, $\sigma$
    Ori, and the ONC are shown as dashed lines for reference.
    \label{cepOB3bage_2}}
\end{figure}

We use the spline fit to the data (Section
\ref{fitting}), overlaid on the fiducial isochrones from Section
\ref{fids} to age the subgroup. This reveals an interesting result.
The subgroup Cep OB3b can clearly be seen to lie on the same locus as
NGC2264, with $\sigma$ Ori being slightly older, implying an age for
Cep OB3b of $\simeq3$ Myrs (Figure \ref{cepOB3bage}). The best previous
age estimate for this subgroup is \cite{1996A&A...312..499J} at 5.5
Myrs.  The older estimate would have made Cep OB3b the oldest group to contain molecular
material. We therefore conclude that the lower age of 3 Myrs is more
likely. The distance modulus uncertainty range is $\pm0.20$ mags. The
effect of this can be seen in Figure \ref{cepOB3bage_2}, where the
relative position of Cep OB3b does not change dramatically even when
the entire uncertainty budget is used.  In addition the photometry
shown in Figure \ref{cepOB3b1} reveals evidence for an R-C gap at the
head of the PMS in a position similar to that of NGC2264 at an age of
$\simeq$3 Myrs, also suggesting a similar age for the two groups.

\subsection{IC348}
\label{IC348_discuss}
\begin{figure}
  \vspace*{174pt}
  \includegraphics{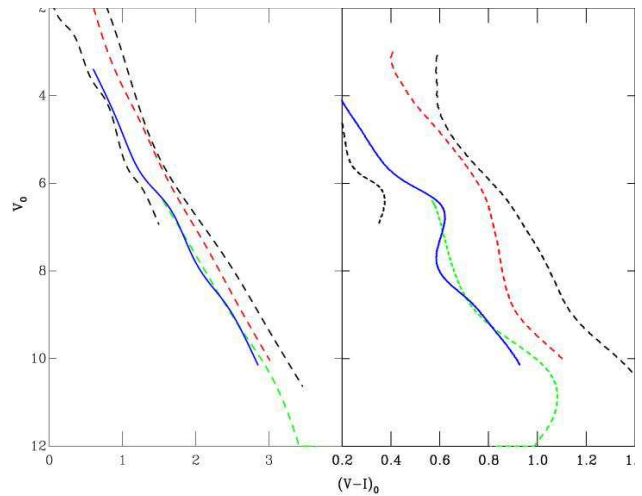}
  \caption{The fitted sequence for IC348 (solid line). h and $\chi$ Per, $\sigma$ Ori, NGC2264 and
    the ONC are shown as dashed lines for reference. 
   \label{ic348age}}
\end{figure}

Using our relative age system, as in Section \ref{cepOB3b_discuss} we
found IC348 to have an age between that of $\sigma$ Ori and NGC2264.
The empirical isochrone therefore suggests that the cluster is
somewhat older than NGC2264 and indeed Cep OB3b. This can be seen in
Figure \ref{ic348age}.  This contrasts with the literature age of 2-3
Myrs \citep[e.g.][]{2001ApJ...553L.153H}.  To obtain an age of 3 Myrs
or less would require a shift of $\simeq0.5$ mags in the distance
modulus, yet the uncertainties in the distance modulus to IC348 are
significantly smaller than this, $-0.16/+0.14$ mag (See Appendix
\ref{appendix2}).

\subsection{$\lambda$ Ori}
\begin{figure}
  \vspace*{174pt}
  \includegraphics{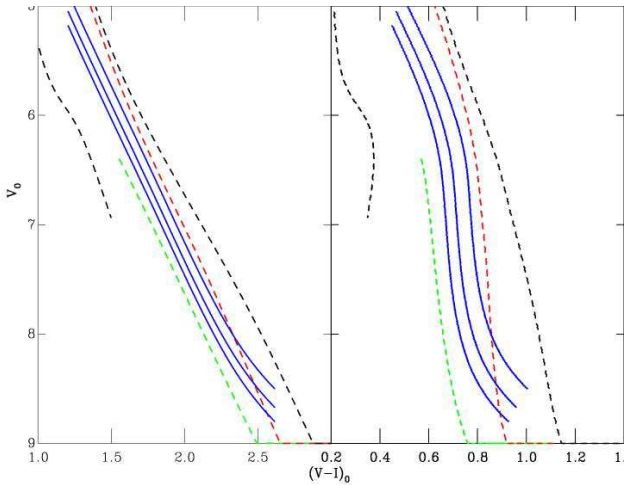}
  \caption{The sequence for $\lambda$ Ori assuming the distance moduli from
    \citet{2001AJ....121.2124D}, \citet{1977MNRAS.181..657M} and
    \textit{HIPPARCOS} (solid lines).  The ONC, NGC2264, $\sigma$ Ori
    and h and $\chi$ Per are shown as dashed lines for reference.
    \label{lambdaori_age}}
\end{figure}

Figure \ref{lambdaori_age} shows the fiducial sequence plot with
$\lambda$ Ori placed at distance moduli of 7.77, 7.90 and 8.07 mags,
representing the limits of the \textit{HIPPARCOS} measurement.  Figure
\ref{lambdaori_age} shows $\lambda$ Ori is largely degenerate with
NGC2264, and therefore has an age of $\simeq3$ Myrs. 
This is certainly consistent with the MS turn-off age of 4 Myr derived from 
UBV photometry by \cite{1977MNRAS.181..657M} but not with the age of 
6-7 Myr derived by \cite{2001AJ....121.2124D} using narrow band
photometry of high mass stars).

\subsection{NGC2362}
\begin{figure}
  \vspace*{174pt}
  \includegraphics{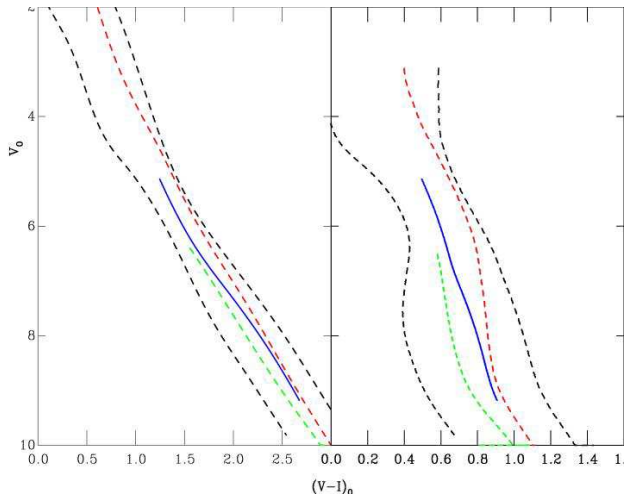}
  \caption{The fitted sequence for NGC2362 (solid line). 
    The ONC, NGC2264, $\sigma$ Ori and h and $\chi$
    Per are shown as dashed lines for reference.
    \label{ngc2362_age}}
\end{figure}

The relative age plot can be seen in Figure \ref{ngc2362_age}. Here
NGC2362 appears between NGC2264 and $\sigma$ Ori, giving it an age of
3-4 Myrs.  This is consistent with the ages derived by
\cite{2005AJ....130.1805D} of $\simeq$1.8 Myrs using the isochrones of
\cite{1997MmSAI..68..807D}, and 3.5-5 Myrs from the isochrones of
\cite{1998A&A...337..403B}.  However, the shift in distance modulus to
achieve an age of greater than 4 Myrs would be $\simeq0.4$ mags;
larger than the typical distance modulus uncertainty.

\subsection{IC5146} 
\begin{figure}
  \vspace*{174pt}
  \includegraphics{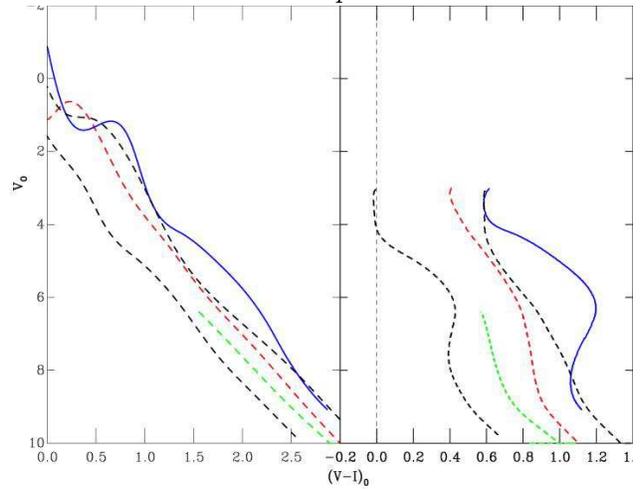}
  \caption{The fitted sequence for IC5146 (solid line), assuming the
    distance modulus from \citet{2002AJ....123..304H}.  The ONC,
    NGC2264, $\sigma$ Ori and h and $\chi$ Per are shown as dashed
    lines for reference. The vertical dashed line shows the
    position of the ZAMS \label{ic5146_age}}
\end{figure}

The age plot is displayed as Figure \ref{ic5146_age}. IC5146 appears
to have an age approximately equal to that of the ONC.  The median age
found by \cite{2002AJ....123..304H} is $\simeq$ 1 Myr, agreeing with
our result.

\subsection{NGC6530}
\begin{figure}
  \vspace*{174pt}
  \includegraphics{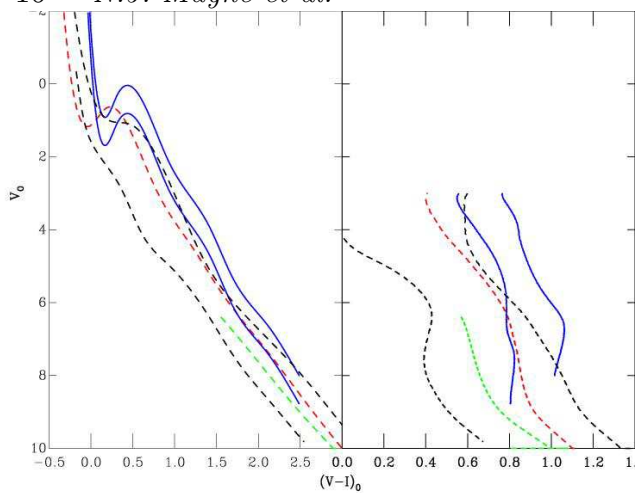}
  \caption{The fitted sequence for NGC6530 (solid line). 
    The ONC, NGC2264, $\sigma$ Ori and h and $\chi$ Per are shown as dashed
    lines for reference. 
    \label{ngc6530_age}}
\end{figure}

Photometry from this cluster is taken from \cite{2005A&A...430..941P}.
They only provide photometry of stars which are also in the catalogues
of \cite{1957ApJ...125..636W} or \cite{1977MNRAS.178..423K} or matched
to an X-ray source from \cite{2004ApJ...608..781D}, and are brighter
than $V=18$.  \cite{2005A&A...430..941P} use the isochrones of
\cite{2000A&A...358..593S} fitted to PMS stars to obtain a median age
of 2.3 Myrs.  \cite{2000AJ....120..333S} find an age of 1.5 Myrs from
fitting isochrones to an H$\alpha$ selected PMS after conversion to an
H-R diagram. \cite{2004ApJ...608..781D}, using X-ray selection and the
photometry of \cite{2000AJ....120..333S} find an age of 0.5-1.5 Myrs.
As for previous fields this spline has been overlaid on the fiducial
diagram seen as Figure \ref{ngc6530_age}.  Two distance moduli are
displayed, that of \cite{2005A&A...430..941P} ($d_m\simeq10.48$ mags)
and that of \cite{2000AJ....120..333S} ($d_m=11.25\pm0.1$ mags).  The
sequence in Figure \ref{ngc6530_age} appears to lie above the ONC for
the greater distance modulus and coincident, or perhaps just below,
the ONC, for the lower distance modulus. Suggesting an age of $<1$ Myr
at the further distance or $\simeq1-2$ Myrs for the closer distance.
Thus we cannot differentiate between the literature ages.

\subsection{NGC7160 and NGC2547}
These clusters have older age estimates than the fields aged using our
earlier fiducial selection (Section \ref{fids}). NGC7160 has an age
estimate of $\simeq10$ Myrs from isochrone fitting from
\cite{2005AJ....130..188S}. NGC2547 has age estimates varying from
30-45 Myrs from MS isochrone fitting in \cite{2002MNRAS.335..291N} and
the Lithium Depletion Boundary in \cite{2005MNRAS.358...13J}.
Therefore we plot these empirical isochrones with h and $\chi$ Per as
our only useful fiducial. NGC7160 is clearly younger than h and $\chi$
Per, making the current age estimates reasonable.  NGC2547 cuts across
the other sequences at redder \textit{V-I}. This could be a problem
with the standard calibrations, which extends only to a \textit{V-I}
of 2.5 in the NGC2547 reduction.

\begin{figure}
  \vspace*{174pt}
  \includegraphics{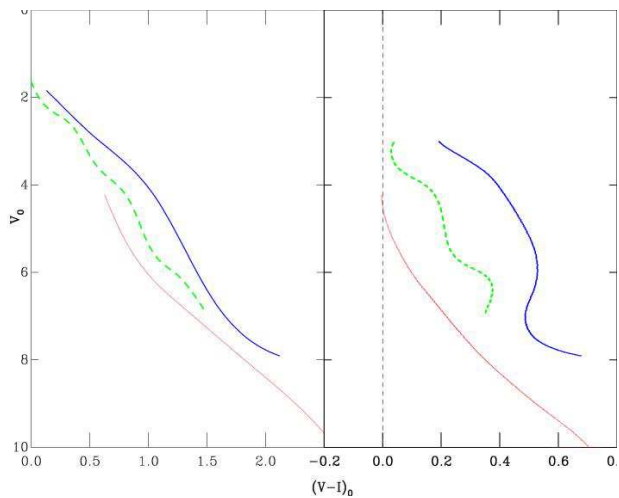}
  \caption{The fitted sequences (solid lines) for (younger to
    older); NGC7160 (blue, $\simeq10$) and NGC2547 (red,$\simeq30-45$
    Myrs) . The dashed line is the spline curve for h and $\chi$ Per
    ($\simeq13$ Myrs). The vertical dashed line shows the position
    of the ZAMS \label{old}}
\end{figure}

\section{Implications of the revised ages}

The consequences of our revision of the relative ages of IC348,
NGC2264, NGC2362 and the ONC are particularly pertinent to recent
studies in two areas.
Firstly in gyrochronology.
Large-sample rotation studies now exist for a subset
of young clusters and from these a conventional view of the 
rotational evolution of stars between 1 and 5 Myr is developing.
The rotational period distribution of the ONC is strongly bi-modal. 
This distribution is believed to evolve into a uni-modal distribution 
such as that of NGC2264 and NGC2362. 
IC348 has a distribution of rotation rates similar to that of the ONC,
and yet we find it to be older than NGC2264.  
This clearly raises a concern for the universality of any evolutionary
model.

\begin{figure}
  \vspace*{174pt}
  \includegraphics{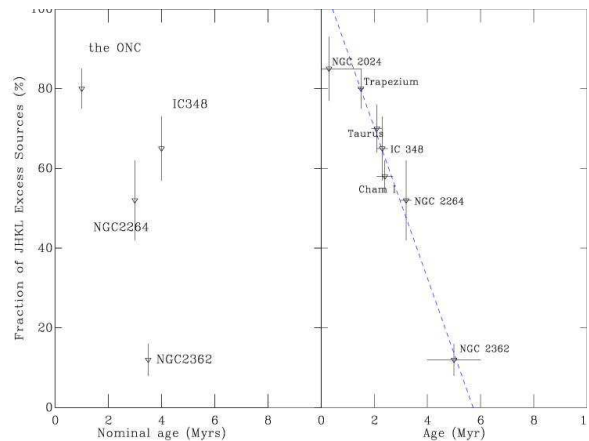}
  \caption{Fraction of JHKL excess sources, used a disc indicator for
    the ONC, IC348, NGC2264 and NGC2362 from
    \citet{2001ApJ...553L.153H}; \textit{left} panel, against nominal
    ages found in this work, \textit{right} panel, against cluster
    ages used in \citet{2001ApJ...553L.153H}. The straight line is a
    simple linear regression fit.\label{disc_new}}
\end{figure}

The second area our findings affect is the field of disc evolution. 
\cite{2001ApJ...553L.153H} plot the fraction of JHKL excess sources,
used to infer a disc fraction as a function of age
for different clusters or groups. 
This plot reveals a clear decrease in the inferred disc fraction with time. 
However, from this work two clusters,
IC348 and NGC2264 should be relocated within this plot leading to a
change in the finer structure of the trend at younger ages.  The disc
fractions for the ONC, IC348, NGC2264 and NGC2362 are $\simeq80\%$,
$\simeq65\%$, $\simeq52\%$ and $\simeq12\%$ respectively. The
relative age order of these four clusters found in this work (youngest
to oldest) is the ONC, NGC2264, NGC2362 and IC348.  Therefore they do
not show a secular decline in disc fraction. This is illustrated
most clearly by the Figure \ref{disc_new}, showing
the disc fraction as a function age from \cite{2001ApJ...553L.153H}
and as a function of the nominal ages found in this work.

Interestingly, IC348 is out of place in rotation rate and disc
evolution.
If disc evolution drives the rotational evolution, say through disc
locking, this is perhaps unsurprising.
The key here may be the local environment.  
There are no O-stars associated with IC348. NGC2362 however has 6 
O-stars stars designated as cluster members, the ONC has 16 and
NGC2264 13 \citep{2004ApJS..151..103M}. 
This supports the
idea that a larger fraction of the PMS stars in IC348 have retained
discs due to the absence of winds and/or ionising radiation from
massive stars.
Our relative ages may, therefore, be amongst the first indications of the
importance of environment on disc and rotation rate evolution
\citep[see also][]{2004AJ....128..765S}.

An alternative explanation has been put forward in
\cite{2003MNRAS.338..616L} by indentifying NGC2362 as anomolous.
\cite{2003MNRAS.338..616L} suggest this could be due to an incorrect
age determination; this is ruled out is by this work. They also
suggest it could be an artifact of the poor limiting magnitude
achieved when using the \textit{L} band, meaning this band is very
sensitive to distance and provides a disc fraction derived from a
different mass range for clusters at an unequal distance. Therefore
any conclusions reached on the secular trends of disc fractions must
be tentative, due to the heterogenous nature of the stellar mass range
used.

\section{Conclusions}
\label{conclusion}
We have presented photometry for a large selection of young stellar
populations in differing environments. 
The selection criteria and some
of the photometry is heterogeneous and from many literature sources.
Despite these difficulties this attempt at producing a self-consistent
set of empirical isochrones has been largely successful. 
We have characterised and fitted the sequences in CMD space, and
grouped the sequences by position in the CMD -- effectively age.
Then using the clearest sequences
typifying each group, a relative age ladder was developed for use with
additional sequences. 
Our final groupings for all the sequences we have data for were as follows.

\begin{enumerate}
\item $\simeq$1 Myr; The ONC, NGC6530 and IC5146.
\item $\simeq$3 Myrs; Cep OB3b, NGC2362, $\lambda$ Ori and NGC2264.
\item $\simeq$4-5 Myrs; $\sigma$ Ori and IC348.
\item $\simeq$10 Myrs; NGC7160-uncertain due to lack of members.
\item $\simeq$13 Myrs; h and $\chi$ Per.
\item $\simeq>$30 Myrs; NGC2547
\end{enumerate}

The distance moduli remain the largest source of uncertainty, allowing the
possibility of changing a sequence by plus or minus one group in age.
Cep OB3b is thus somewhat younger than previously thought, and IC348 
somewhat older.
Both the disc fraction and rotation-period distribution for IC348
imply it is younger than our derived age.
Since, in contrast to the other regions we have studied IC348 has no
O-stars, this suggests that the winds and/or ionising radiation from
these stars have a significant effect of accretion disc evolution.

The data also allow us to follow the feature in the CMD we term the R-C gap. 
In most CMDs this feature appears as a gap between the PMS and MS members, or
as a dearth of members at the head of the PMS in some younger
sequences. We have been able to trace the evolution of the gap from
1 Myr through to the intermediate age (in the context of these data) clusters of h
and $\chi$ Per. 
We have explored the physics behind the
appearance of a gap in the \textit{V}, \textit{V-I} sequences and
the evolution of the gap with age. 
It is encouraging that the size of the gap, where
it is visible, follows the same age ordering as found using the
empirical isochrones. 
Further to this we have touched
upon the possibility of using the size of the R-C gap as a 
distance and extinction independent age indicator. 

The R-C gap could also be used as a powerful tool to calculate age
spreads within a sequence. If one could account for all possible forms
of spread within a CMD, and had memberships and spectral types/masses
for all stars at the boundaries of the R-C gap. By finding the lowest
mass star to have already traversed the R-C gap, and then finding the
youngest star (via isochrones) of the same mass on the PMS one could
calculate an age spread.

Within this study is a limited investigation of the fit across
sequences of theoretical isochrones. We conclude that isochrone
formulations are self consistent when applied within a restricted mass
or colour range not bridging the R-C gap. They tend to become less
reliable when we have attempted to fit both the MS and PMS i.e. both
sides of the R-C gap, excluding the intermediate age clusters, h and
$\chi$ Per. Most recent studies in the literature concentrate on
either the turn-on region of a sequence or the low-mass sections,
meaning that the inconsistency of the fits is not always apparent.

Finally, we tentatively suggest a critical change in the nature of
spreads within a CMD at $\simeq$5 Myrs. Above $\simeq$5 Myrs
the sequence members are generally not associated with molecular
material and have wholly explicable photometric spreads. Below
$\simeq$5 Myrs the sequence members are often associated with
molecular material and show inexplicably large photometric spreads.

To revisit these or other clusters with a larger photometric study
applying the same mechanisms as in this paper would be beneficial and
we believe could provide a benchmark collection of empirical
isochrones for comparison to theory. In unison with this a
comprehensive study of the application and limitations of the R-C gap
as an age indicator would be a logical next step.

\section[]{ACKNOWLEDGMENTS}
NJM is funded by a UK particle physics and astronomy research council
(PPARC) studentship. The INT is operated on the island of La Palma by
the Isaac Newton Group in the Spanish Observatorio del Roque de los
Muchachos of the Instituto de Astrofisica de Canarias. This
publication makes use of data products from the Two Micron All Sky
Survey, which is a joint project of the University of Massachusetts
and the Infrared Processing and Analysis Center/California Institute
of Technology, funded by the National Aeronautics and Space
Administration and the National Science Foundation. We would also like
to thank S.Dahm and G.Herbig for the provision of extra data.

\bibliographystyle{mn2e}
\bibliography{references}
\appendix
\section{Literature memberships}
\label{Known_memberships}
This section details the memberships for each field. Table
\ref{Sources} shows the group, the data type and the reference for
each of the groups in this section.

\subsection{NGC2264}
\label{NGC2264_fit}
\begin{figure}
  \vspace*{174pt}
  \includegraphics{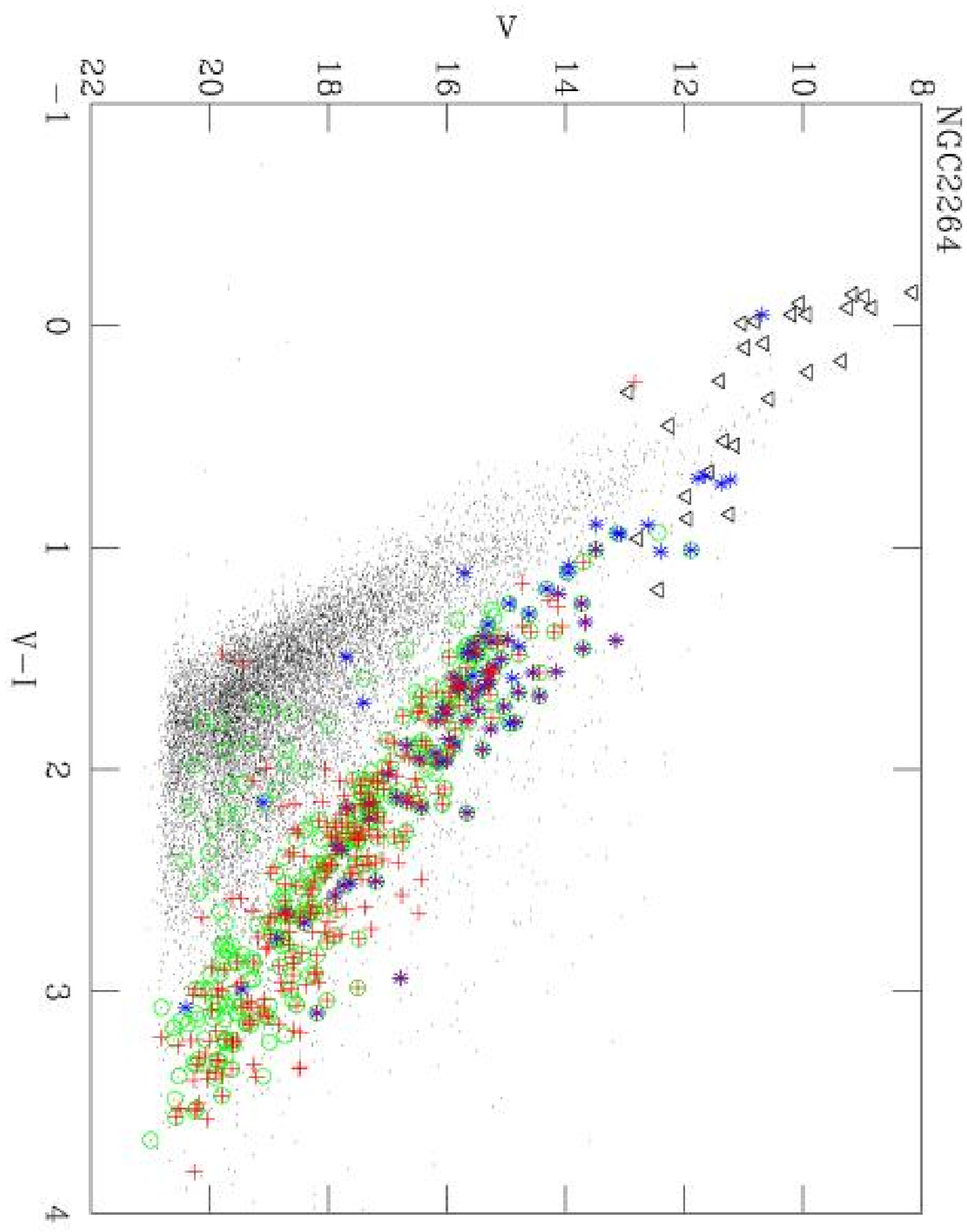}
  \caption{The full catalogue for the NGC2264 field (dots). 
  Circles are the periodic variables from
  \citet{2004A&A...417..557L}, asterisks are X-ray sources from
  \citet{1999A&A...345..521F} and crosses are H$\alpha$ sources from
  \citet{2005AJ....129..829D}. 
  \label{ngc22641}}
\end{figure}

The X-ray sources (Table 10) picked out a reasonably clear sequence
(see Figure \ref{ngc22641}), although there were also some X-ray
active foreground and background objects.  The spectroscopic members
(Table 11) came from \cite{2005AJ....129..829D} who selected members
using H$\alpha$ strength from narrow band imaging, a selection
mechanism which is unbiased in colour-magnitude space. Spectroscopic
observations were taken as follow up. The periodic variables (Table 9)
came from \cite{2004A&A...417..557L} and \cite{2005A&A...430.1005L}.
We have used the entire periodic variable catalogue, whereas the final
membership list of \cite{2005A&A...430.1005L} imposed further
isochrone driven selection. Our photometry has been supplemented
with the photometry for bright stars from \cite{1980MNRAS.190..623M},
as these stars were saturated in our catalogue. 
We include only those stars with probability of membership greater
than 90 percent from the proper motions of
\citep{1980MNRAS.190..623M}.
We have also excluded W33; the anomalous photometry of this
star was also noted in \cite{1980MNRAS.190..623M}. The full CMD with
each form of membership criterion added is shown here as Figure
\ref{ngc22641}.

\subsection{NGC2547}
\label{ngc2547mem}
\cite{2005MNRAS.358...13J} carried out spectroscopy (Table 5) of a
photometrically selected sample. The photometric sequence was clear of
the contamination in this instance and so the colour-magnitude
selection should not have produced a bias in the CMD space. Radial
velocity (RV) was used to define cluster memberships.  The presence of
Li below the Li boundary was also used. The X-ray members (Table 4)
are from \cite{1998MNRAS.300..331J}. 

\subsection{ONC}
\label{onc_mem}
\cite{1997AJ....113.1733H} calculated extinctions for a subset of
stars, using spectroscopic measurements of stars already identified as
members from a range of literature sources, in a spatially selected
region encompassing the inner 15' radius. We have used only those
stars with spectroscopically calculated extinctions with assigned
membership probabilities above 80\%.  Periodic variables were from
\cite{2002A&A...396..513H} and X-ray sources from
\cite{2003ApJ...582..398F}.

\subsection{NGC7160}
\label{ngc7160mem}
Members were taken from \cite{2004AJ....128..805S} and
\cite{2005AJ....130..188S} (Table 7). In these studies a photometric
sample of stars was selected, initially by considering those lying
above the 100 Myr isochrone in a \textit{V}, \textit{V-I} CMD and
lying within a magnitude range of \textit{V}=15-19 mags. A further
selection was then made of stars with \textit{RI} variability, then
using \textit{U}-band and 2MASS colors (biased to accreting stars).
Spectroscopy was then used to measure extinctions and therefore assign
memberships.  The probability of membership was fixed to the distance
in standard deviations ($\sigma$) from the average extinction value
for the group.  This relies on intrinsic colours. The members selected
are those within 1$\sigma$ of the average extinction.

\subsection{$\sigma$ Ori}
\begin{figure}
  \vspace*{174pt}
  \includegraphics{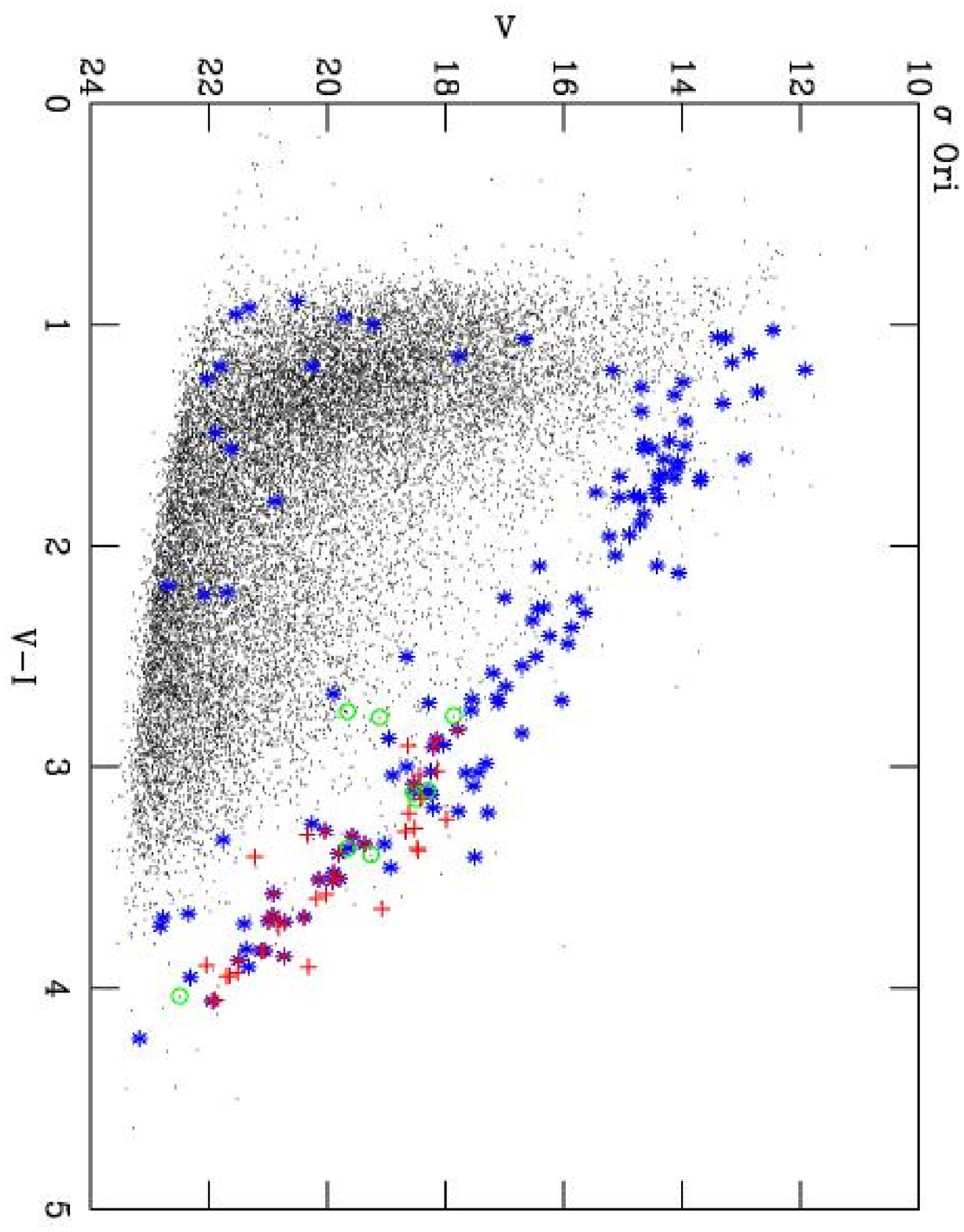}
  \caption{The full catalogue for the $\sigma$ Ori field (dots). 
    Circles are members from \citet{2005MNRAS.356.1583B},
    while asterisks are X-ray sources from
    \citet{2004A&A...421..715S}.  Crosses are members from
    \citet{2005MNRAS.356...89K}. 
    \label{sori1}}
\end{figure}

Spectroscopic members were taken from \cite{2005MNRAS.356...89K} and
\cite{2005MNRAS.356.1583B}. \cite{2005MNRAS.356...89K} obtained
spectroscopy of a photometrically selected sample using a ``close''
selection around the suspected sequence in a CMD, which is potentially
photometrically biased. Memberships were then confirmed on the basis
of measurements of RV and EWs (Equivalent Widths) of NaI consistent
with the group mean. \cite{2005MNRAS.356.1583B} however, used a
``broad'' colour-magnitude selection. The memberships were also
confirmed here via RV and NaI and LiI EWs. \cite{2005MNRAS.356.1583B}
showed that using a ``broad'' selection does not reveal significantly
more members, implying that using members from
\cite{2005MNRAS.356...89K} does not significantly bias our results. Of
the members from \cite{2005MNRAS.356...89K}, only those satisfying all
the criteria were included. In the case of \cite{2005MNRAS.356.1583B}
a selection of $>80\%$ probability of membership was applied to ensure
a high percentage of genuine group members (Table 21). X-ray positions
were taken from \cite{2004A&A...421..715S}. We included those for
which \cite{2004A&A...421..715S} (Table 20) found no optical
counterpart and then cross-correlated the positions with our
catalogue.  The \cite{2004A&A...421..715S} positions were used in
preference to ROSAT data due to the higher spatial resolution of
XMM-Newton. \cite{2006MNRAS.tmpL..62J} shows that PMS members of this
association are separated into two kinematic subgroups by heliocentric
radial velocities, having different ages. In this work all but four
stars from the sequence members are at a declination of less than
$\delta$=-2 18' 00.0''. This area is dominated by members from group 2
as defined in \cite{2006MNRAS.tmpL..62J}. The resulting CMD is
displayed as Figure \ref{sori1}.

\subsection{Cep OB3b}
\label{CepOB3b_members}
Spectroscopic members were taken from \cite{2003MNRAS.341..805P}. They
used a broad colour-magnitude selection with subsequent spectroscopic
measurements of LiI EW, H$\alpha$ and RV.  \cite{2003MNRAS.341..805P}
then used the spectroscopic measurements in conjunction with X-ray
measurements to compare each star to the group mean and assign
memberships (Table 14). H$\alpha$ memberships were taken from
\cite{2002AJ....123.2597O} (Table 17), who used a photometrically
unbiased narrow band imaging survey. The X-ray data were taken from
\cite{1999MNRAS.302..714N}, supplemented with the second ROSAT PSPC
catalogue. The latter contained two pointings; the one from
\cite{1999MNRAS.302..714N}, and a second, non-overlapping field. Both
the unpublished and \cite{1999MNRAS.302..714N} catalogues were
therefore used (Table 14), but where possible we took the positions
from the \cite{1999MNRAS.302..714N} reduction.  In addition the X-ray
sources from \cite{2006ApJS..163..306G} have been included (Table 18).
These sources are from the \textit{CHANDRA} \textit{ACIS} detector,
and subsequently have a much higher spatial resolution and sensitivity
than ROSAT. Both the ROSAT and \textit{CHANDRA} data have been used as
the latter is centralised on a smaller field of view about the cluster
core, while many of the ROSAT detections are outside this area. In the
cases where a \cite{2006ApJS..163..306G} source lies within the
positional error box of the sources from the ROSAT catalogues, the
\textit{CHANDRA} \textit{ACIS} positions have been used. Periodic
variables come from Littlefair et al (in preparation) (Table 16).

\begin{figure}
  \vspace*{174pt}
  \includegraphics{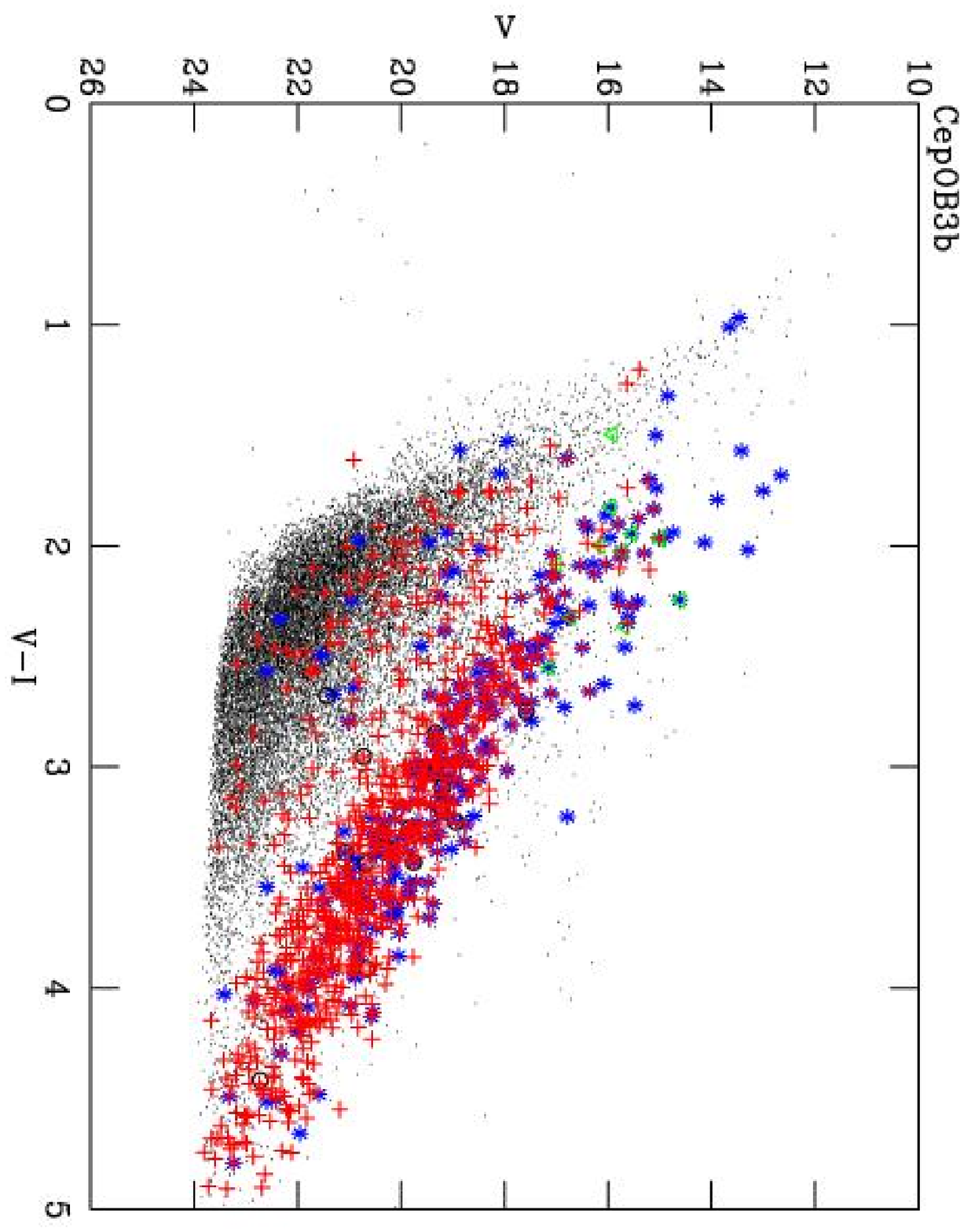}
  \caption{
    The full catalogue for the Cep OB3b field (dots). 
    Asterisks are X-ray sources from \citet{1999MNRAS.302..714N},
    \citet{2006ApJS..163..306G} and the Second ROSAT PSPC catalogue.
    Triangles are members from \citet{2003MNRAS.341..805P}. 
    Circles are H$\alpha$ sources from \citet{2002AJ....123.2597O}. 
    Crosses are the periodic variables from Littlefair et al (in
    preparation).
    \label{cepOB3b1}
    }
\end{figure}

\subsection{IC348}
\label{ic348_members}
\begin{figure}
  \vspace*{174pt}
  \includegraphics{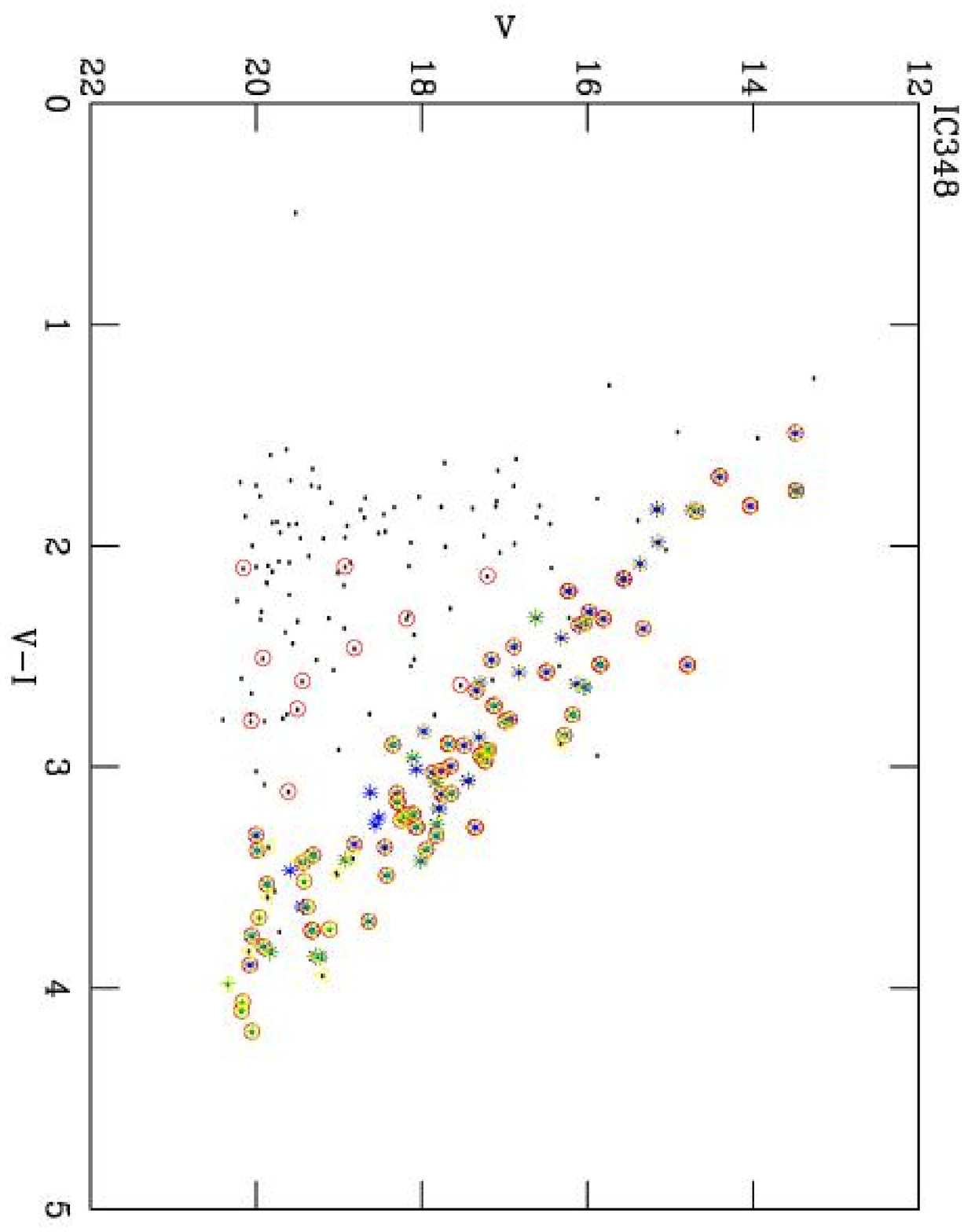}
  \caption{The full catalogue for the IC348 field (dots) 
    The asterisks are X-ray sources from
    \citet{2002AJ....123.1613P} and the Second ROSAT PSPC catalogue.
    Circles are the periodic variables from
    \citet{2004AJ....127.1602C} and \citet{2005MNRAS.358..341L}.
    Crosses are H$\alpha$ sources from \citet{1998ApJ...497..736H}.
    Triangles are spectroscopic members with extinctions from
    \citet{2003ApJ...593.1093L} and \citet{1998ApJ...497..736H}.
    \label{ic3481}}
\end{figure}

Spectroscopic and H$\alpha$ data were taken from \cite{1998ApJ...497..736H}. 
He used photometry
and a wide field H$\alpha$ survey (grism spectrograph), discarding
H$\alpha$ EWs $<2$\AA (Table 27). Follow-up spectroscopy was
undertaken on 80 stars within the field. \cite{1998ApJ...497..736H}
classified the spectra by comparison with dwarf spectral standards,
using these to derive extinctions (Table 26). The second set of
spectroscopic members were taken from \cite{2003ApJ...593.1093L}
(supplemented by \cite{1999ApJ...525..466L},
\cite{2005ApJ...618..810L}, \cite{2005ApJ...623.1141L} and
\cite{2003ApJ...593.1093L}) (Table 26). In \cite{2003ApJ...593.1093L}
spectral types were assigned using various models and then extinctions
calculated by deredenning the stars onto standard colours. The
individual extinctions from \cite{1998ApJ...497..736H} and
\cite{2003ApJ...593.1093L} are used here. X-ray members were taken
from \cite{2002AJ....123.1613P} (Table 24), where H$\alpha$ and
\textit{CHANDRA} data were presented with extinctions included.
\cite{2002AJ....123.1613P} also include extinctions from a range of
literature sources. We supplemented the X-ray data from
\cite{2002AJ....123.1613P} with a portion of the Second ROSAT PSPC
catalogue (Table 25). Periodic variables were taken from
\cite{2004AJ....127.1602C} and \cite{2005MNRAS.358..341L} (Table 23),
both being derived from \textit{I} band wide-field surveys. The
sequence with the members is displayed as Figure \ref{ic3481}.

\subsection{$\lambda$ Ori}
\label{lambda_ori_members}
Likely PMS members were selected from an \textit{R}, \textit{R-I} CMD.
This sample was then spectroscopically observed using Li EW as an
indicator of membership. The have used only those stars showing
significant evidence for Li.

\subsection{NGC2362}
\label{ngc2362_members}
\cite{2005AJ....130.1805D} used a wide field H$\alpha$ survey to
identify a sample of 200 stars, which lay above the ZAMS in an optical
CMD. This sample is then used for follow up spectroscopy, using
H$\alpha$ EW and Li EW as membership criteria. The members so
identified are used here.

\subsection{IC5146}
\label{ic5146_members}
\cite{2002AJ....123..304H} used a wide field survey to identify any
stars showing
$W(H\alpha)>5$\AA. We used stars with known spectral types and exhibiting a
$W(H\alpha)>5$\AA, discarding any stars lying below the Pleiades MS
following \cite{2002AJ....123..304H}.

\subsection{NGC6530}
\label{ngc6530_members}
X-ray members are from \cite{2005A&A...430..941P} using the
\textit{CHANDRA} ACIS detector. H$\alpha$ members are taken from
\cite{2000AJ....120..333S}.

\section{Literature distances and extinctions}
\label{appendix2}
When obtaining parameters from the literature, as far as possible we
avoided using values derived from PMS (Pre-Main Sequence) fitting,
since we wish to derive ages independent of existing PMS isochrones.
However, values derived from MS fitting (i.e fitting the upper part of
the sequence) were used as these are independent of assumptions about
the age. The following section details the literature sources used for
each field.

\subsection{h and $\chi$ Per}
Since the catalogues were combined and normalised to $\chi$ Per, only
the parameters for this cluster were required. The parameters for this
cluster were taken from \cite{2002A&A...394..479C}, $A_V=1.71$,
$d_m=11.7\pm0.1$ mags. This study fitted narrow-band photometry with
MS isochrones. The values of extinction and distance modulus are in
line with other recent determinations such as:
\cite{2001A&A...372..477M} $E(b-y)=0.39\pm0.05$ and
$d_m=11.56\pm0.20$; \cite{2002ApJ...576..880S} $d_m=11.85\pm0.05$ and
$E(\textit{B-V})=0.56\pm0.01$.

\subsection{NGC2547}
\cite{1982A&AS...47..323C} undertook a photoelectric survey in
\textit{UBV} and found 22 probable radial velocity members with
$(\textit{B-V})<0.1$. Using individual reddenings to MS intrinsic
colours he derived $A_V=0.192$, from $E(\textit{B-V})=0.06\pm0.02$
mags, with $d_m=8.27$ mags.  \cite{1999A&A...345..471R} used
\textit{HIPPARCOS} parallax data to calculate a
$d_m=8.18_{-0.26}^{+0.29}$ pc. This later distance modulus used an
alternative method to isochronal or intrinsic colour fitting, so will
be most useful here. The studies used for memberships in Section
\ref{ngc2547mem} have all confirmed within uncertainties the
parameters found in these two cases. The carried forward values are;
$d_m=8.18_{-0.26}^{+0.29}$ mags and $A_V=0.192$.

\subsection{ONC}
\label{onc_parameters}
$\rm{H_20}$ MASER measurements are used in \cite{1981ApJ...244..884G},
where expansion velocities of molecular clouds were used to infer the
distance modulus. A result of $d_m=8.38\pm0.37$ mags was obtained.
This along with extinctions from \cite{1997AJ....113.1733H} were used
for this field.

\subsection{NGC2264}
In \cite{2005AJ....129..829D} a summary of values from previous
studies is presented along with the methods of calculation. All the
entries listed are from forms of isochrone fitting, but we consider
only those studies fitting the main sequence.
\cite{1997AJ....114.2644S} used colour-magnitude selection in
$R-H\alpha$, then fitted the ZAMS to assumed B type stars. An average
distance modulus of $9.4\pm0.25$ mags and $A_V=0.23$ resulted.
\cite{1987PASP...99.1050P} used shifts of individual OB type stars to
intrinsic colours leading to $A_V=0.19$ and $d_m=9.88\pm0.17$ mags.
\cite{2000AJ....120..894P} selected stars using H$\alpha$ and X-ray
emission, then used MS isochrone fitting with the bright stars,
deriving a distance modulus of $9.4\pm0.10$ mags and
$E(\textit{B-V})=0.066\pm0.034$. An average of these values has been
taken being $d_m=9.6$ mags and $A_V=0.21$.
 
\subsection{$\sigma$ Ori}
\cite{1994A&A...289..101B} found parameters using the lower part of
the sequence, without the use of isochrones. Derivations of surface
gravity from narrow band photometry were calibrated to intrinsic
colours of the theoretical and empirical models in
\cite{1981Ap&SS..80..353S}. \cite{1994A&A...289..101B} used MK
spectral types in conjunction with effective temperature and absolute
bolometric magnitudes to calculate surface gravities and radii of
stars. In \cite{1981Ap&SS..80..353S} however the calibrations were
done without taking into account PMS or post MS objects.
\cite{1994A&A...289..101B} applied these calibrations to all spectral
types, clearly at the age of around 3 Myr many of these stars are
indeed PMS objects. A set of subgroups were selected and the subgroup
1b in this paper covers our field. The values for this subgroup were
given as $d_m=7.8\pm0.39$ mags, with a value of
$E(\textit{B-V})=0.06$.  The distance modulus from \textit{HIPPARCOS}
parallax measurements is quoted as $d_m=7.73_{-0.60}^{+0.84}$ to the
star $\sigma$ Orionis. The large uncertainties make this value
unusable.  Therefore the values taken forward were from
\cite{1994A&A...289..101B}, namely $d_m=7.8\pm0.39$ mags and
$E(\textit{B-V})=0.06$ ($A_V=0.18$).

\subsection{NGC7160}
\cite{2004AJ....128..805S} and \cite{2005AJ....130..188S} used
membership criteria described in Section \ref{ngc7160mem}. These
papers used spectroscopically derived spectral types and intrinsic
colours to calculate the extinctions, and MS isochrone fitting for the
distance modulus. The resulting values were $A_V=1.17$ and $d_m=9.77$
mags.

\subsection{Cep OB3b}
\label{cepOB3b_dist}
\cite{1993A&A...273..619M} used optical and IR photometry of O, B and
A type stars. These were compared with MS calibrations to calculate
individual extinctions and distance moduli, which were then averaged.
\cite{2003MNRAS.341..805P} recalculated these parameters from the
\cite{1993A&A...273..619M} individual extinctions and distance moduli,
but excluded star 11 from \cite{1959ApJ...130...69B} which is of
dubious membership. The resulting values which we also adopted were
$d_m=9.65\pm0.20$ and $A_V=2.81\pm0.10$.

\subsection{IC348}
\label{ic348_parameters}
We adopted the distance modulus of \cite{1974PASP...86..798S}, who
used \textit{UBVKL} photometry of 20 stars with \textit{V}$<14$
believed to be cluster members, in conjunction with various literature
spectral types and MK classifications to derive a distance modulus of
$7.5_{-0.16}^{+0.14}$ mags.  This is the same distance modulus as that
derived to the Per OB2 association, $d_m=7.5$ mags. A full discussion
of the distance to IC348 is presented in \cite{1998ApJ...497..736H}.
The extinctions for this field were individual star extinctions and
were from \cite{1998ApJ...497..736H} and \cite{2003ApJ...593.1093L}.

\subsection{$\lambda$ Ori}
\label{lambda_ori_parameters}
We have used the \textit{HIPPARCOS} distance modulus to this cluster,
$d_m=7.90\pm0.17$. The extinction to this cluster is calculated in
\cite{1994ApJS...93..211D} as E(B-V)=0.12, equating to $A_V=0.36$.

\subsection{NGC2362}
\label{ngc2362_parameters}
\cite{1996MNRAS.281.1341B}, using narrow band photometry of B-type
stars derive $A_V=0.31$ and $d_m=10.87\pm0.03$.
\cite{2001ApJ...563L..73M} also used the B-type stars to obtain
$A_V=0.31$ and $d_m=10.85$.  The values adopted here are
$10.87\pm0.03$ and $A_V=0.31$.

\subsection{IC5146}
\label{ic5146_parameters}
\cite{2002AJ....123..304H} derived a distance modulus from a
comparison of the early-type stars with models yielding
$d_m\simeq10.4$.  They then calculated the reddening by assuming their
distance modulus and using the known spectral types of a selection of
38 stars, to yield $A_V=3.0\pm0.2$. We have adopted $d_m\simeq10.4$.
The individual extinctions used for the B type stars to calculate the
distance modulus have been used, with the average extinction being
applied to the remaining stars. Due to the need for individual
extinctions, as with the ONC and IC348 the extinctions were applied
before fitting.

\subsection{NGC6530}
\label{ngc6530_parameters}
\cite{2005A&A...430..941P} used MS fitting of the stars in the blue
envelope.  These are MS stars assumed to be at the cluster distance or
closer.  They calculate $d_m\simeq10.48$ and use the value of the
extinction from \cite{2000AJ....120..333S}.
\cite{2000AJ....120..333S} fitted a ZAMS relation individually to 30
early-type stars, yielding two different average distance moduli,
assuming one was due to binarity they adopted the larger distance
modulus, $d_m=11.25\pm0.1$ mags.  The Q-method is then used for these
early-type stars yielding $A_V\simeq1.09$.  In
\cite{2005A&A...430..941P} a useful summary table of the distance
moduli calculated to this cluster is presented along with a summary of
the literature of the field.  For this study we have used both
$d_m\simeq10.48$, $d_m\simeq11.25$ and used $A_V\simeq1.09$.

\begin{landscape}
\begin{table}
\begin{tabular}{|l|l|l|l|l|l|l|l|l|l|l|l|l|l|l|}
Field \& CCD (field.ccd)&ID&RA&DEC&x position (CCD)&y position
(CCD)&MAG&UNCERTAINTY&FLG&COL&UNCERTAINTY&FLG\\
\hline
1.04&173&02 22 19.173&+57 08
11.19&978.686&2770.194&12.230&0.010&OO&0.386&0.016&OO\\
\hline
\end{tabular}
\caption{The members for each field are available on the cluster
home page and in the electronic version of this paper; this table is a
sample. \label{catalogues}}
\begin{tabular}{|l|l|l||l||l|l||l|l|l||l|}
\hline
Field&$A_V$&Source&Method&$d_m$(mags)&Source&Method&Age(Myrs)&Source\\
\hline
NGC2547&$0.15$&1&Q-method&$8.18_{-0.26}^{+0.29}$&11&HIPPARCOS&30-45&27\\
\hline
$\chi$ Per&$1.57\pm0.079$&2&Q-method&$11.7\pm0.1$&10&MS fitting&12.6&2\\
\hline
NGC2264&$0.371$&4&Q-method&9.6&13&Avg of many values&3&\\
\hline
NGC7160&$1.17$&5&Intrinsic colours&9.77&14&Isochrone fitting&10&5\\
\hline
Cep OB3b&$2.882$&6&Q-method&$9.65\pm0.20$&15&Spectroscopy&5.5&20\\
\hline
$\sigma$ Ori&$0.18$& 8,9&Iso fitting&$7.8\pm0.39$&17&HIPPARCOS&3&17\\
\hline
ONC&$A*$&19&Spectral classes&$8.38\pm0.37$&18&$\rm{H_2O}$
MASER&$\approx0.8$&19\\
\hline
IC348&A*&7&Spectral classes
classification&$7.5_{-0.16}^{+0.14}$&16&HIPPARCOS&2-3&7\\
\hline
$\lambda$ Ori&0.37&24&Spectral classes&$7.90\pm0.17$&24&HIPPARCOS&4,
6-7&28\\
\hline
NGC2362&0.31&25&Average of literature values&$10.87\pm0.03$&26&MS
fitting&$\simeq1.8$, 3.5-5&25\\
\hline
IC5146&$3.0\pm0.2$&23&Spectral type classification&10.4&23&MS
fitting&1&23\\
\hline
NGC6530&1.09&22&MS fitting&10.48&22&MS fitting&2.3, 1.5 or
0.5-1.5&29, 30 and 31\\
\end{tabular}
\caption{Extinction, $d_m$, ages and sources. A* = Individual values from sources indicated. \label{dm_AV}}
\begin{tabular}{|l|l|l|}
\hline
Field&Q-method&Literature\\
$\chi$ and h Per&1.571 and 1.728&1.82 and 1.71\\
NGC2264&0.371&0.22\\
NGC2547&0.15&0.192\\
Cep OB3b&2.882&$2.81\pm0.1$\\
\hline
\end{tabular}
\caption{Extinctions (mags) from Q-method and from the literature. \label{Qt}}
1 \cite{1982A&AS...47..323C}, 2 \cite{2002ApJ...576..880S}, 3 \cite{2000A&A...357..471S}, 4
  \cite{1980MNRAS.190..623M}, 5 \cite{2004AJ....128..805S}, 6
  \cite{1996A&A...312..499J}, 7 \cite{1998ApJ...497..736H} and
  \cite{2003ApJ...593.1093L}, 8 \cite{1999ApJ...521..671B}, 9
  \cite{2004AJ....128.2316S},\cite{1994A&A...289..101B}, 10
  \cite{2002A&A...394..479C}, 11 \cite{1999A&A...345..471R}, 12
  \cite{2000A&A...357..471S}, 13 \cite{2005AJ....129..829D}, 14
  \cite{2004AJ....128..805S}, 15 \cite{2003MNRAS.341..805P}, 16
  \cite{1998ApJ...497..736H}, 17 \cite{2004AJ....128.2316S}, 18
  \cite{1981ApJ...244..884G}, 19 \cite{1997AJ....113.1733H}, 20
  \cite{1996A&A...312..499J}, 21 \cite{1994ApJS...93..211D}, 22
  \cite{2005A&A...430..941P}, 23 \cite{2002AJ....123..304H}, 24
  \cite{1994ApJS...93..211D}, 25 \cite{2005AJ....130.1805D}, 26
  \cite{1996MNRAS.281.1341B}, 27 \cite{2005MNRAS.358...13J}, 28
  \cite{1977MNRAS.181..657M} and \cite{2001AJ....121.2124D}, 29
  \cite{2000A&A...358..593S}, 30 \cite{2000AJ....120..333S} and 31
  \cite{2004ApJ...608..781D}
  The used values of extinction for the lower age sequences is largely
  inconsequential as the isochrones are parallel to the dereddening
  vector at the lower mass end.
\end{table}
\end{landscape}
\bsp
\label{lastpage}
\end{document}